\documentclass{aastex}

\received{}
\accepted{}

\slugcomment{Submitted to ApJ}

\newcommand{\ms}{\mbox{m s$^{-1}~$}}
\newcommand{\msun}{M$_{\odot}$}

\newcommand{\mjup}{M$_{\rm JUP}$}

\newcommand{\msini}{$M \sin i~$}

\shortauthors{Vogt {\it et~al.\/}}
\shorttitle{Six New Keck Planets}
\begin{document}

\title{Six New Planets from the Keck Precision Velocity Survey$~^{1}$}

\author{Steven S. Vogt\altaffilmark{2}, 
Geoffrey W. Marcy\altaffilmark{3},
R. Paul Butler\altaffilmark{4}, 
Kevin Apps\altaffilmark{5}}

\email{vogt@ucolick.org}

\altaffiltext{1}{Based on observations obtained at the W.M. Keck
Observatory, which is operated jointly by the University of California
and the California Institute of Technology. Keck Telescope time comes
from both NASA and the University of California}

\altaffiltext{2}{UCO/Lick Observatory, 
University of California at Santa Cruz, Santa Cruz, CA, USA 95064}

\altaffiltext{3}{ Department of Astronomy, University of California,
Berkeley, CA USA  94720 and at Department of Physics and Astronomy,
San Francisco, CA, USA 94132}

\altaffiltext{4}{Department of Terrestrial Magnetism,
5241 Broad Branch Road NW, Washington D.C. USA 20015-1305}

\altaffiltext{5}{ Physics and Astronomy, University of Sussex,
Falmer, Brighton BN1 9QJ, UK}

\begin{abstract}

We report results of a search for planets around 500 main sequence
stars using the Keck high resolution spectrometer which has provided
Doppler precision of 3 \ms during the past 3 yr.  We report 6 new
strong planet candidates having complete Keplerian orbits, with
periods ranging from 24 d to 3 yr.  We also provide updated orbital
parameters for 4 previously announced planets.

Four of the six newly discovered planets have minimum \msini masses
less than 2 \mjup, while the remaining two have \msini $\sim$ 5 \mjup.
The distribution of planetary masses continues to exhibit a rise
toward lower masses.  The orbital eccentricities of the new planets
range from 0.12 to 0.71 which also continues the ubiquity of high
eccentricities.  All 18 known extrasolar planets orbiting beyond 0.2
AU have eccentricities greater than $\sim$0.1. The current limiting
Doppler precision of the Keck Doppler survey is 3 \ms per observation
as determined from observations of both stable stars and residuals to
Keplerian fits.

\end{abstract}


\section{Introduction}
\label{intro}

To date, 22 planet candidates have been identified around nearby main
sequence stars by measuring their Keplerian Doppler shifts.  Four
groups have contributed the bulk of the detections by surveying a
total of $\sim$300 stars at a Doppler precision of $\sim$10 \ms
(cf. Marcy, Cochran, Mayor 2000, Noyes et al. 1997).  These
``planets'' all have mass estimates, \msini, less than 7.5 \mjup where
$i$ is the unknown orbital inclination.

Fortuitously, these precision Doppler surveys, along with
low--precision surveys of several thousand stars, have revealed only
11 orbiting brown dwarf candidates, \msini = 8--80 \mjup, and most are
actually hydrogen--burning stars with low orbital inclination (Mayor
et al. 1997, Halbwachs et al. 1999, Udry et al. 1999).  This paucity
of brown dwarf companions renders the planet candidates
distinguishable by their high occurrence at low masses: 17 of the 22
have \msini = 0.4--4 \mjup (cf. Figure 6, Butler and Marcy 1997;
Marcy, Cochran \& Mayor 2000).

The planet candidates reveal a mass distribution that rises toward
lower masses, from $\sim$8 \mjup to 0.5 \mjup which is the lowest
\msini currently detected.  Remarkably, all 13 planets that orbit
beyond 0.2 AU reside in non--circular orbits with $e > 0.09$ and many
higher than 0.3.  In contrast, the Earth and giant planets in our
solar system have eccentricity less than 0.05 .  Planet formation
theory is challenged to find robust mechanisms that produce these
observed distributions of mass and orbital eccentricity (cf. Lissauer
1995, Weidenschilling and Marzari 1996, Lin and Ida 1996, Rasio and
Ford 1996).  Further, the half--dozen planets that reside within 0.2
AU offer a challenge to explain their current location (cf. Lin et
al. 1996).

The system of three planetary--mass companions around the main
sequence star, Upsilon Andromedae, opens questions about the ubiquity
of multiple planets and about the formation mechanisms that could
explain multiple jupiter-mass planets within 3 AU.  One wonders if a
jupiter-mass planet within 3 AU is commonly accompanied by additional
giant planets farther out, as demanded by dynamical evolution
scenarios that involve mutual perturbations.  Further Doppler
measurements of existing and future planets can help ascertain the
occurrence and character of multiple planet systems.

The broad goals of the precision Doppler surveys include the
following: 1) detection of several hundred planets, sufficient to
construct statistically-meaningful distributions of planet mass,
eccentricity, and orbital distance; 2) the detection of Jupiter--mass
planets beyond 4 AU to compare with our Jupiter; 3) the
characterization of multiple planet systems; 4) characterization of
planet distributions down to saturn-masses; 5) assessment of
correlations between planets and stellar properties such as
metallicity (i.e. Gonzalez et al. 1999).  Toward achieving these
goals, full-sky surveys of more than 1,000 stars are being carried out
by our group, by Mayor's group (Mayor et al. 1999), and by others.
Most main sequence dwarf stars brighter than V=7.5 are currently being
surveyed, with a need for more surveys in different regimes of
parameter space.

This paper reports the discovery of six new planet candidates from the
Keck extrasolar planet survey.  Section 2 describes the Keck precision
velocity program including technique, the stellar sample, and the
current level of precision.  The stellar properties and Keplerian
orbital fits for the six new planet candidates are presented in
Section 3.  Section 4 provides an update on the orbital parameters for
several previously announced planets.  A discussion follows.

\section{The Keck Planet Search Program}
\label{obs}

The Keck Doppler planet survey began in July 1996 using Keck I with
the HIRES echelle spectrometer (Vogt et al. 1994). The spectra have
resolution, R=80,000 and span wavelengths from 3900--6200 \AA.
Wavelength calibration is carried out by means of an iodine absorption
cell (Marcy \& Butler 1992; Butler et al. 1996) which superimposes a
reference iodine spectrum directly on the stellar spectra.

The stellar sample contains main sequence stars from F7--M5.  Stars
hotter than F7 contain too few spectral features to achieve precision
of 3 \ms, while stars later than M5 are too faint ($V>$11) for the
Keck telescope to achieve 3 \ms precision in our nominal 10-minute
exposure time.  The G \& K dwarfs are mostly within 50 pc and selected
from the Hipparcos catalog (Perryman et al. 1997), while M dwarfs have
been selected from both Hipparcos and the Gliese catalog.  Evolved
stars have been removed from the observing list based on Hipparcos
distances.

The list has been further sieved to remove chromospherically active
stars as these stars show velocity ``jitter'' of 10 to 50 \ms, related
to rapid rotation, spots, and magnetic fields (Saar et al. 1998).  The
Ca II H\&K line reversals are used as a chromospheric diagnostic
(Noyes et al. 1984), and are measured directly from our Keck HIRES
spectra.  The H\&K measurements are placed on the Mt Wilson ``S''
scale by calibration with previously published results (Duncan et
al. 1991; Baliunas et al. 1995; Henry et al. 1996).  Based on their
``S'' index, stars with ages less than 2 Gyr are either excluded from
our sample, or are given shorter exposure times.

Stars with known stellar companions within 2 arc seconds are removed
from the observing list as it is operationally difficult to get an
uncontaminated spectrum of a star with a nearby companion.  Otherwise,
there is no bias against observing multiple stars.  Further, the list
of Keck program stars has no bias against brown dwarf companions.
Stars with known or suspected brown dwarf companions have not been
excluded from the Keck target list.

Doppler measurement errors from our Keck survey were previously
reported to be 6 to 8 \ms (Butler et al. 1998; Marcy et al. 1999).
However, we have since made improvements to the data analysis
software.  We now achieve a precision of 3 \ms with HIRES by treating
readout electronics and CCD charge diffusion in the model that
determines the instrumental PSF of each observation (Valenti et
al. 1995).  We routinely achieve precision of 3 \ms for $V=$8 stars in
10 minute exposures, as shown in Figures 1, 2, and 3, which cover
stars of spectral types late F and G, K, and M respectively. The
observed velocity precision, $\sigma_{obs}$, is the quadrature sum of
the instrumental errors, $\sigma_{inst}$, and random velocity
variations intrinsic to the program stars, $\sigma_{star}$. Saar et
al. (1998) have shown that, for the slowest rotating G, K, and M
stars, $\sigma_{star} \sim 2$ \ms. Thus, our instrumental precision is
presently $\sigma_{inst} \sim 2$ \ms per observation. All previous
HIRES data has also been reprocessed to bring its precision to this
level. With further improvements, we expect to achieve photon limited
precision of $\sim$ 2 \ms with this system, and thus our exposures are
nominally taken at a S/N sufficient for this goal.

\section{New Planet Candidates from the Keck Survey}
\label{news}

Six new planet candidates have recently emerged from the Keck survey.
For HD 192263, the discovery of its planet was announced during the
writing of this paper (Santos et al. 1999).

The stellar properties of the six host stars are given in Table 1.
The first three columns provide the common name, the HD catalog
number, and the Hipparcos catalog number respectively.  Spectral types
are from a calibration of $B-V$ and Hipparcos-derived absolute
magnitudes.  The stellar masses are estimated by interpolation of
evolutionary tracks (Fuhrmann et al. 1998; 1997).  The values of
R'$_{\rm HK}$, a measure of the ratio of chromospheric to bolometric
flux (Noyes et al. 1984), are measured from the CaII H\&K line cores
in the Keck spectra.  Distances are from Hipparcos (Perryman et
al. 1997).

The [Fe/H] values are based on a calibration of the Hauck \&
Mermilliod (1997) catalog of uvby photometry and 60 [Fe/H]
determinations from high resolution spectroscopy (Favata et al. 1997;
Gonzalez et al. 1999; Gonzalez et al. 1998; Gonzalez 1998; Gonzalez
1997). From the scatter in the calibration relationship, the
uncertainty in [Fe/H] is estimated to be 0.07 dex for stars from G0 V
to K0 V.

Astrophysical effects that can mimic Keplerian Doppler velocity
signals include radial and nonradial pulsations and rotational
modulation caused by stellar surface features (i.e. spots, plages).
Five of the new Keck candidates are chromospherically inactive, with
R'$_{\rm HK}$ values similar to the Sun or lower.  In contrast, the
sixth star, HD 192263, is extremely active with R'$_{\rm HK}$ = -4.37
(Santos et al. 1999).

The orbital parameters for the six new Keck planet candidates are
listed in Table 2. The quantities in parentheses are the formal
uncertainties in each orbital parameter, as determined by monte carlo
simulations. The individual Keck Doppler velocity measurements are
listed in Tables 3 through 8. The host stars are discussed below.

\subsection{HD 10697}

HD 10697 (HR 508) is assigned a spectral type of G5 IV by the Bright
Star Catalog (Hoffleit 1982) and Hipparcos. From its spectral type,
$B-V$ color and the Hipparcos--derived absolute magnitude, we find a
stellar mass of 1.10 \msun for HD 10697 based on placement on standard
evolutionary tracks. This star is a slow rotator and is
chromospherically inactive, as indicated by R'$_{\rm HK}$ = -5.02, by
its photometric stability (Lockwood et al. 1997), and by its low X-ray
flux (Hunsch et al. 1998). Relative to the Sun, HD 10697 is modestly
metal rich, [Fe/H]=+0.15 .

We have made 30 velocity observations of HD 10697, spanning three
years, as shown in Figure 4 and listed in Table 3. These observations
barely cover one orbital period of the new planet candidate,
accounting for the relatively large uncertainty in the derived orbital
period. The amplitude ($K$) of the Keplerian orbital fit is 119 \ms,
while the RMS of the velocity residuals to the Keplerian fit is 7.8
\ms. This scatter is twice that of our present known errors, for
reasons not yet known. The eccentricity of the orbit, $e$=0.12, is
about twice as large as that of Jupiter, as is common for extrasolar
planet candidates that orbit beyond 0.2 AU.  Its minimum mass, \msini,
is 6.35 \mjup, placing it among the most massive companions found from
precision velocity surveys.

The semi-major axis of the orbit is $a$ = 2.12 AU, yielding a maximum
angular separation between planet and star of 73 mas.  The amplitude
of the astrometric wobble of the star is 373/{$\sin i~$} $\mu$as,
making this a prime target for interferometric astrometry. The
expected effective temperature of the planet due to stellar insolation
(assuming an albedo of 0.3) is 264K (Saumon et al. 1996), and internal
heating may increase this by 10 - 20 K. Direct detection would be
difficult at present.

\subsection{HD 37124}

HD 37124 (G4V) is a slowly rotating, chromospherically inactive star
with R'$_{\rm HK}$ = -4.90.  As listed in Table 4 and shown in Figure
5, 15 velocity measurements have been made, spanning 2.7 years,
revealing six complete orbits with a derived orbital period of 155.7
d.  The semi-amplitude ($K$) is 43 \ms and the eccentricity is 0.19,
giving the companion a minimum mass of 1.04 \mjup.  The RMS of the
observations to the Keplerian fit is 2.82 \ms

The semi-major axis of the companion orbit is $a$ = 0.55 AU, yielding a
maximum angular separation between planet and star of 20 mas. The
planet is expected to have $T_{\rm eff}$ = 327K (Saumon et al. 1996).
HD 37124 has low metallicity, [Fe/H]=-0.32, relative to the Sun, but
its metallicity is nearly typical for field GK dwarfs. This is the
lowest metallicity star known to have a planet. Most previous planet
candidates have been found around metal rich host stars.

\subsection{HD 134987}

HD 134987 (HR 5657, G5V) is similar to 51 Pegasi in its spectral type,
enhanced metallicity, and low chromospheric activity.  But its planet
has $P$=259 d and \msini = 1.58 \mjup.  We have made 43 Doppler
observations spanning 3 years, as shown in Figure 6 and listed in
Table 5.  The eccentricity is 0.24, quite non-circular as with all
extrasolar planet candidates orbiting beyond 0.2 AU.  The RMS of the
Keplerian fit to the measured velocities is 3.0 \ms, consistent with
measurement errors.
 
The semi-major axis of the planetary orbit is $a$ = 0.81 AU, yielding a
maximum angular separation between planet and star of 39 mas.  We
expect a planet temperature of $T_{\rm eff}$=315K (Saumon et
al. 1996). The amplitude of the associated astrometric wobble is
45/{$\sin i~$} $\mu$as.

\subsection{HD 177830}

HD 177830 is an evolved subgiant of spectral type K2IV, with
photometry placing it close to giant status, $M_{\rm V}$=3.32. It is
difficult to estimate [Fe/H] of subgiants from photometry and we are
not aware of a spectroscopic analysis of metallicity. The stellar mass
estimate of 1.15$\pm$0.2 \msun is based on placement on evolutionary
tracks but is similarly suspect.  Figure 7 shows a spectroscopic
comparison of HD 177830 (dotted line) and the chromospherically
inactive K dwarf $\sigma$ Dra (light solid line) in the core of the Ca
II H line.  The heavy solid line is the chromospherically active K2 V
star HD 192263 which will be discussed in the next subsection.  Even
slowly rotating, chromospherically inactive, main sequence K dwarf
stars, like $\sigma$ Dra, show mild core reversal in Ca II H \& K
lines, while evolved K subgiants, like HD 177830, have ``flat
bottomed'' line cores.

We have made 29 velocity measurements of HD 177830 spanning 3 yr, as
shown in Figure 8 and listed in Table 6.  The orbital fit yields
$P$=391.6 d, $K$=34 \ms, and $e$=0.41, yielding \msini = 1.22 \mjup.
The RMS of the velocity residuals to the Keplerian fit is 5.2 \ms,
again larger than our known errors.

The somewhat elevated RMS of 5.2 \ms is probably related to the
subgiant status of HD 177830. This star is a very evolved subgiant,
with M$_v$ = 3.32 $\pm$ 0.1. True K III giants often show complex
velocity variability with periods ranging from less than a day to
several hundred days.  $\beta$ Oph, a K2 III giant, shows evidence of
multiple periodicities with times scales of less than 1 d and velocity
amplitudes of $\sim$40 \ms (Hatzes \& Cochran 1994).  Velocity
variations with periods of several hundred days have been observed in
$\pi$ Her (Hatzes \& Cochran 1999), Aldebaran and $\beta$ Gem (Hatzes
\& Cochran 1993).  Arcturus (Hatzes \& Cochran 1993; 1994) shows both
short and long period Doppler velocity variations with amplitudes of
order 150 \ms.  In contrast Horner (1996) surveyed 4 late--type giants
with spectral types ranging from G8 III to K2 III and found these
stars to be stable at the 25 \ms level.

Little is known about the intrinsic velocity stability of subgiants.
The three stars on the Keck precision velocity program which most
closely match the $B-V$ and absolute magnitude of HD 177830 are HD
136442, HD 208801, and HD 23249.  These stars are stable at the 4 to 6
\ms level. It is thus probable that the observed velocity variations
of HD 177830 are due to Keplerian orbital motion, but we cannot rule
out intrinsic photospheric variations.  This star should be followed
up with precision photometry and line bisector studies.

The semi-major axis of the planetary orbit is $a$ = 1.10 AU, yielding a
maximum separation between planet and star of 27 mas. The planet is
expected to have $T_{\rm eff}$=362K (Saumon et al. 1996), with some
increase due to internal heating.

\subsection{HD 192263}

A planet around this star was recently announced by Santos et
al. (1999) during the writing of this paper. They found the following
orbital parameters: $P$=23.87$\pm$0.14 d, $K$=65 \ms, and a small but
uncertain orbital eccentricity.  Here we find: $P$=24.4$\pm$0.07 d,
and $K$=68$\pm$11 \ms, as listed in Table 2. Our 15 Keck velocity
measurements are shown in Figure 9 and listed in Table 7. The
eccentricity, 0.03, derived from 40 observations of Santos et al. is
somewhat lower than that derived from our 15 Keck observations,
$e$=0.22$\pm$0.14, which indicates that the eccentricity is small but
requires further observations to establish firmly. We find that
\msini=0.78 \mjup, $a$=0.15 AU, implying an expected $T_{\rm eff}$=486
K. (Saumon et al. 1996).

HD 192263 is a rapidly rotating, chromospherically active, K0 dwarf
with R'$_{\rm HK}$=-4.37 (Santos et al. 1999).  Figure 7 shows the
line core reversal of the Ca II H line for HD 192263 (dark solid
line).  The slowly rotating chromospherically inactive K dwarf
$\sigma$ Dra is shown for comparison.  Based on the measured R'$_{\rm
HK}$ value of -4.37, we estimate that the intrinsic Doppler ``jitter''
for this star is 10 to 30 \ms (Saar et al. 1998).  Figure 10 shows the
periodogram of the chromospheric ``S'' value measurements from our 15
Keck spectra.  The highest peak corresponds to a period of 26.7 days,
uncomfortably close to the observed Doppler velocity period of 24 d.
The false alarm probability for this periodogram peak is 4.7\%,
suggesting the peak may be coincidental. It remains possible that the
velocity periodicity is not due to an orbiting body, but rather to
surface effects.  However, the stellar rotation period is expected to
be $\sim$8 d from the chromospheric strength. This short rotation
period argues against rotational modulation of surface features as the
cause of the velocity variations having $P$=24 d.

Hipparcos found photometric variability for HD 192263 of $\sim$0.011
mag. which is low given its high chromospheric activity.  This star
should be subject to intensive photometric monitoring.  If photometry
or ``S'' value measurements continue to show periodicities similar to
the observed Doppler velocity period, this would suggest that the
source of the variations is intrinsic to the star rather than to an
orbiting planet

The rotational Doppler broadening in the Keck/HIRES spectra implies
$V\sin i <$ 3.0 km s$^{-1}$, consistent with that found, $V\sin i =$
1.8 $\pm$1.2 km s$^{-1}$, by Santos et al. (1999).  Given the short
rotation period of $\sim$8 d, implied by the chromospheric activity,
along with low $V\sin i$, the star must be viewed within 30 deg of
pole--on.  Such a pole--on vantage point is consistent with the nearly
constant photometric brightness reported by Hipparcos.  Apparently the
chromospheric activity and spotted regions remain visible on the
(polar) hemisphere during an entire rotation period.  If the orbital
plane of the companion is also nearly pole-on, its mass is
considerably higher than \msini of 0.78 \mjup.  We also remain puzzled
by the RMS of 4.5 \ms of the velocity residuals to our Keplerian fit.
This RMS is just too low for such an active star, which should exhibit
jitter of at least 10 \ms . Perhaps it is because the inclination
angle is so low, and thus there is not much radial velocity jitter
produced by the surface brightness inhomogeneities since they aren't
strongly rotationally modulated. We will continue to monitor the ``S''
value and Doppler velocities. For now, we are not yet completely
convinced of a planet-companion interpretation for the velocity
variations of HD 192263.

\subsection{HD 222582}

HD 222582 is a near solar ``twin'' in spectral type, metallicity, and
chromospheric activity (Table 1).  Figure 11 shows a comparison of the
solar flux spectrum (Kurucz et al. 1984) and a spectrum of HD 222582,
centered on the core of the Ca II H line with a resolution of
R$\approx$80,000.  The many narrow absorption lines due to neutral
metal atoms (mostly Fe I) have the same widths and depths in HD 222582
and the Sun, indicating that the two stars have similar temperature,
abundances, and $V\sin i$.  The core of the H line, formed in the
chromosphere, is deeper in HD 222582, suggesting slightly weaker
chromospheric and magnetic activity than the Sun.

We have made 24 Doppler observations spanning 1.7 years, as shown in
Figure 12 and listed in Table 8. Aliasing allows two possible
Keplerian solutions. The preferred Keplerian, shown in Figure 12, has
$P$=575.9 d, a velocity semi--amplitude, $K$=184 \ms, and an
eccentricity, $e$=0.71, yielding a minimum mass, \msini=5.29 \mjup.
The RMS to the Keplerian orbital fit is 3.4 \ms. The second possible
orbital fit has $P$=529 d and an RMS of 6.5 \ms. This orbital
ambiguity exists because the observed time span of the observations is
only slightly longer than the orbital period.  We expect to resolve
this within the next 6 months.
 
The semi-major axis of the orbit is $a$ = 1.35 AU, yielding a maximum
angular separation between planet and star of 55 mas. The amplitude of
the astrometric wobble is 165/{$\sin i~$} $\mu$as. The effective
temperature of the planet is expected to be 234K (Saumon et al. 1996).

\section{Update of Previously Announced Planets}

Orbital parameters for several previously announced extrasolar planet
candidates are updated in this section based on recent Keck
observations.  The stellar properties of these stars are listed in
Table 9, which has the same format as Table 1.  The updated orbital
parameters are listed in Table 10, while the individual Keck Doppler
velocity measurements are listed in Tables 11 through 14 for HD
187123, HD 195019, HD 210277 and HD 217107, respectively.

We cannot at this time update the orbits for two stars, HD 168443
(Marcy et al. 1999) and GL 876 (Marcy et al. 1998, Delfosse et
al. 1998), for which planetary companions were found previously in the
Keck Doppler survey.  As noted in those discovery papers, both stars
continue to show small secular departures from simple Keplerian
motion, making the fit to a single orbit imprecise. In the case of HD
168443, the velocity residuals to the Keplerian fit now exhibit a
longterm trend with significant curvature (as of October, 1999),
strongly indicative of a second companion with long period. However,
more data are required to test the Keplerian nature of the velocity
residuals for both stars and to place constraints on plausible orbits.

\subsection{HD 187123}

The Keck velocities for HD 187123 are listed in Table 11, and the
phased velocities are shown in Figure 13 with a linear trend
removed. The updated orbital elements shown in Table 10 are consistent
with the discovery data (Butler et al. 1998), though the RMS to the
Keplerian fit has improved by about a factor of two.

Two years of additional monitoring reveal a linear trend in the
velocities with a slope of -7$\pm$2 \ms per year.  This suggests the
existence of a second companion with a period much longer than 3 yr.
Another few years of monitoring will be required to determine if this
slope is real.  HD 187123 does not have any known companions.

\subsection{HD 195019}

The discovery of the planet around HD 195019 was initially announced
from data collected with both the Lick 3-m and the Keck telescopes
(Fischer et al. 1999). The newly derived orbit given here is in good
agreement with the originally published orbit, but again with smaller
velocity residuals due to the improvements in our Doppler technique.
The measured velocities are listed in Table 12 and the phased
velocities are shown on Figure 14.

With an orbital period of 18.2 d and a semi-major axis of 0.13 AU, this
is the most distant planet yet found in a circular orbit,
$e$=0.02$\pm$0.02.

\subsection{HD 210277}

We have made 45 velocity measurements of HD 210277, listed in Table 13
and shown in Figure 15. The newly derived orbit confirms the orbit
from the discovery paper (Marcy et al. 1999), but the RMS (3.3 \ms) of
the residuals to the orbital fit is again smaller by a factor of
$\sim$2 due to our improved Doppler precision.

The new orbit implies $a$=1.10 AU and $e$=0.45$\pm$0.03.  The maximum
separation between the companion and the primary is 75 mas, and the
amplitude of the astrometric variation is 65/{$\sin i~$} $\mu$as.  The
expected equilibrium temperature of the companion is 243K.

\subsection{HD 217107}

We have made 21 Keck velocity measurements of HD 217107, listed in
Table 14 and shown in Figure 16 with a linear trend of 40 \ms removed
from this plot. The updated orbit agrees with the orbit from the
discovery paper (Fischer et al. 1999), which was based on data from
both the Lick and Keck Observatories.

The current Lick observations (51 measurements) reveal a linear trend
in the velocity residuals to the orbital fit, strongly indicative of a
second companion.  A simultaneous fit of the Lick velocities to a
model composed of a Keplerian orbit plus a trend yields a best--fit
trend of 43.3$\pm$2.8 \ms per yr and a period of $P$=7.127$\pm$0.001 d
(Fischer 1999, private comm.).  We attempted the same type of model
fit to the Keck velocities and found a best--fit trend of 39.4$\pm$3.5
\ms per yr and $P$=7.126$\pm$0.001 d.  Thus the velocity trend in the
Lick data are confirmed independently by those from Keck.

The Keck data yield an orbital eccentricity, $e$=0.14$\pm$0.02, in
agreement with that from Lick (Fischer et al. 1999).  Thus the
non--circular orbit for such a close companion ($a$=0.072 AU) raises
questions about the tidal circularization time scale.  Perhaps the
eccentricity is driven by another companion, indeed possibly that
which causes the linear velocity trend.  The velocities indicate that
any second companion must have a period longer than 2 yr, and a mass
greater than 4 \mjup.  Follow-up work with astrometry and
high--resolution IR imaging is warranted to detect this possible
second companion.

\section{Discussion}

Our Keck program has gathered velocity measurements for 3 years,
making it sensitive to planets orbiting out to 2 AU. Including the 6
new planet candidates described in this paper, the Keck survey has
resulted in the discovery or co-discovery of 12 extrasolar planet
candidates. The current precision of the Keck survey, held over the
3-year time base, is 3 \ms, sufficient to eventually make 3$\sigma$
detections of Jupiter mass companions in 5 AU orbits, should they
exist. However, another decade of data will be required before the
Keck survey will begin probing planets in these 5 AU orbits. We are
working to improve single-shot precision with the Keck system to 2
\ms.

Four of the six newly discovered planets have minimum masses (\msini)
less than 2 \mjup. Figure 17 shows the latest mass distribution of the
know extrasolar planets (Marcy, Cochran, and Mayor, 2000). The high
incidence of companions having \msini$<$2 \mjup\ adds further support
to a planetary mass distribution that begins a dramatic rise at about
5-6\mjup, and increases toward the lowest detectable masses from
8\mjup\ to 0.5 \mjup\ (Butler and Marcy 1997). Below 0.5\mjup,
detectability drops markedly. The highest planetary masses are
apparently $\sim$8 \mjup\, which constitutes a physical upper limit
that should be explained by planet--formation theory, perhaps via
tidal truncation of growth in the protoplanetary disk (Bryden et
al. 1999, Nelson et al. 1999).

It is important to keep in mind here that there is no bias against
brown dwarf companions in the Lick, Keck or Geneva surveys. Indeed,
the CORAVEL survey found 11 brown dwarf ``candidates'' orbiting within
5 AU (Duquennoy \& Mayor 1991, Mayor et al. 1997).  However, Hipparcos
astrometry (Halbwachs et al. 1999) has subsequently revealed that most
of these 11 brown dwarf candidates previously identified from the low
precision CORAVEL survey are, in fact, M dwarfs viewed at low
inclination angles.

Nonetheless, companions of 10--80 \mjup\ are much easier to detect
than those of Jupiter--mass.  But the most massive companion found to
date from any precision velocity survey is 70 Vir b, with 7.4 \mjup\
(Marcy \& Butler 1996). Our Keck survey has {\it not} revealed any
``brown dwarf candidates'' (defined by \msini = 10--80 \mjup) out of
the 500 stars being surveyed. Thus the occurrence of brown dwarf
companions within 2 AU resides below 0.5\% in agreement with the few
detections from previous surveys (Cumming et al. 1999, Halbwachs et
al. 1999).

More than 1,000 stars have been surveyed for two years or longer by
the Geneva group and by our surveys, from which five ``51 Peg--like''
planets have emerged, with orbital periods less than 5 days.
Precision Doppler programs are strongly biased toward detecting these
planets. Thus we estimate that $\sim$0.5\% of main sequence stars have
these ``hot jupiter'' companions. Four planets have been found with
orbital periods between 7 and 20 days, and the distribution of orbital
radii is steadily filling in, suggesting a continuous and nearly flat
distribution of semi-major axes above 0.04 AU.

The eccentricities of the six newly announced planet candidates range
from 0.12 to 0.71. All of the extrasolar planets orbiting beyond 0.2
AU, now numbering 18, have eccentricities, $e \ge$ 0.1. For
comparison, the eccentricities of Jupiter and Earth are 0.05 and 0.03,
respectively. HD 222582 has the largest eccentricity, 0.71, of any
planet found to date. Unlike the other extreme eccentricity case, 16
Cyg B, HD 222582 has no known stellar companion.

These eccentric extrasolar planet orbits may arise from gravitational
interactions with other planets or stars, or from resonant interaction
with the protoplanetary disk. Regarding the latter, consider the
planets around HD 177830 and HD 10697. HD 177830 is a case with
relatively low mass and high eccentricity (\msini = 1.22\mjup\ and $e$
= 0.41), while HD 10697 is a case with high mass and low eccentricity
(\msini = 6.35 \mjup\, $e$ = 0.12). Taken together, this pair would
seem to disfavor the theory of eccentricity pumping of planetary
orbits through resonant interaction with the disk (Artymowicz,
1992). That theory predicts that it is the highest mass companions
which become the most eccentric, while the lowest mass companions
actually lose eccentricity. Clearly though, it is still a case of
small number statistics, and many more systems will need to be found
before such trends can be established robustly. Whatever the ultimate
cause of eccentric orbits, the growing ubiquity of high-eccentricity
systems seems to suggest that ``minimum entropy'' planetary systems,
like our own, with its suite of nested coplanar nearly-circular
orbits, may be rare.

Most of the stars with known planetary companions are metal rich
relative to the Sun (Gonzalez et al. 1999; Gonzalez et al. 1998;
Gonzalez 1998; Gonzalez 1997). Two of the six new candidates described
here are metal rich, but two are metal poor (Table 1), one has nearly
solar abundances, and the metallicity of the remaining star is not yet
known. Thus these six new planet--bearing stars do not add any
additional support to the suspected metallicity correlation.

The Keck survey, after 3 years, is now detecting planets out to 2.1
A.U. and is thus marching through and able to probe the interesting
``habitable zone'' (HZ) defined by Kasting et al. (1993) around main
sequence stars. Five of the six new planets lie either directly in, or
near the edges of the habitable zones for their stars. HD 10697 orbits
at 1.87 to 2.39 AU, just at the outer edge of the HZ for a G5V star,
but probably well within the HZ for its evolved G5IV star. Its
equilibrium insolation temperature is expected to be 274-284K. HD
37124 orbits at 0.45 to 0.65 A.U. with an expected temperature of
$T_{\rm eff}$= 327K. It lies at the inner edge of the HZ for its G4V
star. HD 134987 orbits at 0.62 to 1.00 A.U. with a $T_{\rm eff}$=
315K, solidly in the HZ for its G5V star. HD 177830 orbits at 0.63 to
1.57 A.U., with an expected $T_{\rm eff}$= 362K. It would lie at the
inner edge of the HZ for a K2 star, but is probably slightly outside
the inner edge since its star is an evolved subgiant. HD 222582 has an
orbital semi-major axis of 1.35 A.U., but the orbit is quite eccentric
and its orbital distance varies from 0.39 to 2.31 A.U. It has an
expected $T_{\rm eff}$= 234K, and also lies directly in the HZ of its
G3V star. Whether or not water could exist in liquid form, either in
the atmospheres of these gas giants, or possibly on accompanying
moons, is beyond our ability to say at present and requires further
detailed theoretical modeling of planetary atmospheres (Burrows \&
Sharp, 1999).

Complementary extrasolar planet detection techniques include
photometric transit surveys, interferometric astrometry, IR imaging of
dust disks (Trilling et al. 1998, Trilling et al. 1999, Koerner et
al. 1998, Schneider et al. 1999), and spectroscopic searches for
reflected light (Cameron et al. 1999, Charbonneau et al. 1999).
Several of the new planet candidates provide good targets for these
new techniques.  In particular, HD 10697 and HD 222582 will have
minimum astrometric amplitudes of 373 and 165 $\mu$as, and astrometric
detection will yield unambiguous masses.

\acknowledgements

We gratefully acknowledge support by NASA grant NAG5-8861 and NSF
grant AST95-20443 (to GWM), by NSF grant AST-9619418 and NASA grant
NAG5-4445 (to SSV), travel support from the Carnegie Institution of
Washington (to RPB), and by Sun Microsystems. We thank the NASA and UC
Telescope assignment committees for allocations of telescope time. We
also thank Debra Fischer and Doug Lin for useful discussions. This
research has made use of the Simbad database, operated at CDS,
Strasbourg, France.

\clearpage
\begin{figure}
\caption{Keck/HIRES velocities of a representative subset of F and G dwarfs 
from our planet search, chosen to be chromospherically quiet (age $>$ 3 Gyr).
The typical velocity scatter is 3 \ms, which represents the sum
of measurement errors and intrinsic photospheric variability.
The observations span the 2--3 yr duration of our Keck planet search.}
\label{FGdwarfs}
\end{figure}

\begin{figure}
\caption{Velocities of a representative set of K dwarfs, showing typical
scatter of 3--4 \ms that represent errors and intrinsic variability}
\label{Kdwarfs}
\end{figure}

\begin{figure}
\caption{Velocities of a representative set of M dwarfs, showing typical
scatter of 3--4 \ms that represent errors and intrinsic variability.}
\label{Mdwarfs}
\end{figure}

\begin{figure}
\caption{Velocities for HD 10697 (G5 IV) spanning 3 yr.
The solid line is a Keplerian orbital fit with
orbital parameters, $P$=1072 d, $K$=119 \ms,
and $e$=0.12, yielding a minimum mass,
\msini = 6.35 \mjup \ for the companion.  The
RMS of the residuals to the Keplerian fit is 7.8 \ms}

\label{fig4}
\end{figure}

\begin{figure}
\caption{Velocities for HD 37124 (G4 V).
The solid line is a Keplerian orbital fit, giving
$P$=155.7 d, $K$=43 \ms,
and $e$=0.19, yielding 
\msini = 1.04 \mjup.  The
RMS of the residuals 2.82 \ms.}
\label{fig5}
\end{figure}

\begin{figure}
\caption{Velocities for HD 134987 (G5 V).
The solid line is a Keplerian orbital fit, giving
$P$=260 d, $K$=50 \ms,
and $e$=0.24, yielding 
\msini = 1.58 \mjup.  The
RMS of the residuals 3.0 \ms.}
\label{fig6}
\end{figure}

\begin{figure}
\caption{Comparison of Ca II H line core of HD 177830 (dotted line)
with a chromopsherically quiet K2 V star $\sigma$ Draconis (light
solid line).  Slowly rotating K dwarfs show a mild core reversal,
unlike the more evolved subgiants such as HD 177830 which have
flat--bottom cores.  Dramatic line core reversal is seen in the
rapidly rotating, chromospherically active, K2 V star HD 192263 (heavy
solid line).}
\label{fig7}
\end{figure}

\begin{figure}
\caption{Velocities for HD 177830 (K2 IV).  The solid line is a
Keplerian orbital fit with $P$=391.6 d, $K$=34 \ms, and $e$=0.41,
yielding \msini = 1.22 \mjup .  The RMS of residuals is 5.2 \ms.}
\label{fig8}
\end{figure}

\begin{figure}
\caption{Velocities for HD 192263 (K2 V).
The solid line is a Keplerian orbital fit with a
$P$=24.36 days, $K$=68 \ms,
and $e$=0.22, yielding a minimum
\msini = 0.78 \mjup .  The
RMS of the residuals is 4.5 \ms.}
\label{fig9}
\end{figure}

\begin{figure}
\caption{Periodogram of the chromospheric S measurements
for HD 192263.  The primary peak of 26.7 days is similar
to the period, $P$=24 d, exhibited in the velocities, leaving an
ambiguous interpretation of the velocities. The orbital interpretation
is favored (see text).}
\label{fig10}
\end{figure}

\begin{figure}
\caption{Comparison of Ca II H line core for HD 222582 (dotted
line) with the Sun (solid line).  The solar spectrum is remarkably
similar to HD 222582 throughout the visible region.}
\label{fig11}
\end{figure}

\begin{figure}
\caption{Velocities for HD 222582 (G3 V).
The solid line is a Keplerian orbital fit with 
$P$=575.9 d, $K$=184 \ms,
and $e$=0.71, yielding
\msini= 5.29 \mjup .  The
RMS of the residuals is 3.36 \ms, consistent with 
measurement errors.}
\label{fig12}
\end{figure}

\begin{figure}
\caption{Phased velocities for HD 187123 (G3 V).
The solid line is a Keplerian orbital fit with 
$P$=3.096 days, a semi--amplitude of 68 \ms,
and an eccentricity of 0.01, yielding 
\msini = 0.48 \mjup .  The
RMS of the residuals to the Keplerian fit is 3.29 \ms.
A linear trend of -7$\pm$2 \ms per year has been
removed here. This linear trend suggests presence of a second companion
with an orbital period much longer than 2 years.}
\label{fig13}
\end{figure}

\begin{figure}
\caption{Phased velocities for HD 195109 (G3 V).
The solid line is a Keplerian orbital fit with
$P$=18.20 d, $K$=272 \ms,
and $e$=0.02, yielding \msini = 3.47 \mjup .  
The RMS of the residuals to the Keplerian fit is 3.46 \ms.}
\label{fig14}
\end{figure}

\begin{figure}
\caption{Doppler velocities for HD 210277 (G7 V).  The implied orbital
parameters are, $P$=436.6 d, $K$=39 \ms, and $e$=0.45, giving \msini =
1.23 \mjup .  The RMS of the residuals is 3.24 \ms.}
\label{fig15}
\end{figure}

\begin{figure}
\caption{Phased velocities for HD 217107 (G7 V).
The solid line is a Keplerian orbital fit giving,
$P$=7.126 d, $K$=140 \ms,
and $e$ = 0.14, yielding 
\msini = 1.27 \mjup.  
An apparent linear trend of 40 \ms per year has been
removed from the measured velocities prior to performing the fit.  
This linear trend suggests a second companion
with an orbital period much longer than 2 years and mass $>$ 4 \mjup.
The RMS of the residuals to the Keplerian fit is 4.20 \ms.}
\label{fig16}
\end{figure}

\begin{figure}
\caption{Mass histogram for all known companions to main sequence stars}
\label{fig17}
\end{figure}

\clearpage

\begin{deluxetable}{lrrrlllll}
\tablenum{1}
\tablecaption{Stellar Properties}
\label{candid}
\tablewidth{0pt}
\tablehead{
\colhead{Star}  & \colhead{Star} & \colhead{Star} & \colhead{Spec} & \colhead{M$_{\rm Star}$} & \colhead{V} & \colhead{R'$_{\rm HK}$} & \colhead{[Fe/H]} & \colhead{d} \\
\colhead{ }     & \colhead{(HD)} & \colhead{(Hipp)} &\colhead{type} & \colhead{(M$_{\odot}$)} & \colhead{(mag)} & \colhead{ }         & \colhead{ }      & \colhead{(pc)}
} 
\startdata
109 Psc  &  10697 &   8159 & G5 IV & 1.10 & 6.27 & -5.02 & +0.15 & 32.6 \\
         &  37124 &  26381 & G4 ~V & 0.91 & 7.68 & -4.90 & -0.32 & 33.2 \\
 23 Lib  & 134987 &  74500 & G5 ~V & 1.05 & 6.47 & -5.01 & +0.23 & 25.6 \\
         & 177830 &  93746 & K2 IV & 1.15 & 7.18 & -5.28 &       & 59.0 \\
         & 192263 &  99711 & K2 ~V & 0.75 & 7.79 & -4.37 & -0.20 & 19.9 \\
         & 222582 & 116906 & G3 ~V & 1.00 & 7.68 & -5.00 & -0.01 & 41.9 \\
\enddata
\end{deluxetable}

\clearpage

\begin{deluxetable}{rrllllllll}
\tablenum{2}
\tablecaption{Orbital Parameters}
\label{candid}
\tablewidth{0pt}
\tablehead{
\colhead{Star}  & \colhead{Period} & \colhead{$K$} & \colhead{$e$} & \colhead{$\omega$} & \colhead{$T_0$} & \colhead{\msini} & \colhead{a} & \colhead{N} & \colhead{RMS} 
\\
\colhead{(HD)} & \colhead{(days)} & \colhead{(\ms)} &\colhead{ } & \colhead{(deg)} & \colhead{(JD-2450000)}  & \colhead{(\mjup)} & \colhead{(AU)} & \colhead{obs } & \colhead{(\ms)}
} 
\startdata
 10697 &  1072.3 (9.6) & 119 (3) & 0.12 (0.02) & 113 (14) & 1482 (39) & 6.35 & 2.12 & 30 & 7.75 \\
 37124 &   155.7 (2.0) & ~43 (7) & 0.19 (0.12) & ~73 (52) & 1219 (33) & 1.04 & 0.55 & 15 & 2.82 \\
134987 &   259.6 (1.1) & ~50 (1) & 0.24 (0.03) & 353 (10) & 1364 (6) & 1.58 & 0.81 & 43 & 2.99 \\
177830 &   391.6 (11) & ~34 (14) & 0.41 (0.13) & 5 (35) & 1333 (14) & 1.22 & 1.10 & 29 & 5.18 \\
192263 &  24.36 (0.07) & ~68 (11) & 0.22 (0.14) & 153 (18) & 1380.9 (1) & 0.78 & 0.15 & 15 & 4.52 \\
222582 &  575.9 (3.6) & 184 (52) & 0.71 (0.05) & 306 (12) & 1310 (17) & 5.29 & 1.35 & 24 & 3.36 \\
\enddata
\end{deluxetable}

\clearpage

\begin{deluxetable}{rrr}
\tablenum{3}
\tablecaption{Velocities for HD 10697}
\label{vel10697}
\tablewidth{0pt}
\tablehead{
JD           & RV            & error \\
(-2450000)   &  (m s$^{-1}$) & (m s$^{-1}$) 
}
\startdata
\tableline
   367.0617  & -83.1  &  3.1 \\
   715.0682  & -143.7 &  3.1 \\
   716.0976  & -155.3 &  3.0 \\
   983.1289  &   0.0  &  2.8 \\
  1013.0972  &   7.8  &  3.1 \\
  1014.0916  &  11.0  &  2.7 \\
  1043.0700  &  25.5  &  3.4 \\
  1044.0879  &  18.8  &  2.9 \\
  1051.0464  &  25.1  &  2.8 \\
  1068.9138  &  38.2  &  3.0 \\
  1070.0916  &  36.4  &  2.9 \\
  1070.9763  &  36.9  &  3.0 \\
  1071.9868  &  36.8  &  3.1 \\
  1072.9530  &  32.7  &  3.4 \\
  1074.8834  &  29.8  &  2.8 \\
  1075.8794  &  37.0  &  2.7 \\
  1170.8261  &  55.5  &  4.2 \\
  1342.1172  &   7.9  &  3.1 \\
  1343.1171  &  10.0  &  2.8 \\
  1368.1185  & -14.8  &  3.1 \\
  1369.0795  &  -8.3  &  3.3 \\
  1374.1286  & -18.6  &  3.4 \\
  1410.1192  & -44.3  &  2.9 \\
  1411.0250  & -47.6  &  3.3 \\
  1412.0736  & -54.4  &  3.1 \\
  1438.8873  & -76.7  &  3.1 \\
  1439.9187  & -73.3  &  3.0 \\
  1440.9339  & -67.7  &  2.9 \\
  1487.9490  & -112.2 &  3.7 \\
  1488.8072  & -89.5  &  3.7 \\
\enddata
\end{deluxetable}

\clearpage

\begin{deluxetable}{rrr}
\tablenum{4}
\tablecaption{Velocities for HD 37124}
\label{vel37124}
\tablewidth{0pt}
\tablehead{
JD & RV & error \\
(-2450000)   &  (m s$^{-1}$) & (m s$^{-1}$) 
}
\startdata
\tableline
   420.0466  &  47.8  &  3.7 \\
   546.7365  &  30.7  &  2.8 \\
   837.7662  &   1.2  &  2.8 \\
   838.9487  &   4.0  &  3.1 \\
   861.8046  &  23.7  &  3.3 \\
  1069.0362  &  -1.7  &  2.7 \\
  1070.1319  &   0.7  &  2.6 \\
  1071.1149  &   0.0  &  3.3 \\
  1072.1295  &  -4.2  &  3.0 \\
  1073.0296  &  -8.0  &  3.1 \\
  1172.8957  &  32.6  &  3.3 \\
  1226.7806  &  -2.3  &  3.1 \\
  1227.7817  &  -6.0  &  3.2 \\
  1228.7429  & -10.8  &  2.8 \\
  1412.1416  & -38.0  &  4.3 \\
\enddata
\end{deluxetable}

\clearpage

\begin{deluxetable}{rrr}
\tablenum{5}
\tablecaption{Velocities for HD 134987}
\label{vel134987}
\tablewidth{0pt}
\tablehead{
JD & RV & error \\
(-2450000)   &  (m s$^{-1}$) & (m s$^{-1}$) 
}
\startdata
\tableline
   276.8020  & -18.0  &  3.3 \\
   604.8935  &  28.7  &  3.9 \\
   838.1755  &  40.1  &  2.6 \\
   839.1727  &  41.5  &  2.7 \\
   840.1707  &  42.3  &  2.6 \\
   863.1203  &  36.3  &  2.8 \\
   954.9159  & -54.4  &  1.8 \\
   956.9547  & -54.4  &  3.0 \\
   981.8126  & -59.9  &  3.0 \\
   982.8190  & -54.6  &  2.5 \\
   983.8498  & -50.9  &  3.5 \\
  1011.8011  & -51.8  &  2.6 \\
  1012.8005  & -50.7  &  2.7 \\
  1013.8006  & -54.4  &  2.8 \\
  1050.7730  & -16.8  &  3.0 \\
  1051.7547  & -14.0  &  2.6 \\
  1068.7306  &   1.7  &  2.5 \\
  1069.7193  &   4.6  &  2.2 \\
  1070.7242  &   5.1  &  2.2 \\
  1071.7229  &   1.6  &  2.1 \\
  1072.7204  &  11.0  &  2.6 \\
  1073.7196  &   7.2  &  3.4 \\
  1074.7070  &  13.0  &  2.4 \\
  1200.1581  & -47.0  &  2.8 \\
  1227.0883  & -52.9  &  2.7 \\
  1228.1038  & -53.1  &  2.5 \\
  1229.1161  & -51.6  &  2.6 \\
  1310.8892  & -22.7  &  2.8 \\
  1311.9101  & -20.3  &  2.8 \\
  1312.9239  & -19.2  &  3.1 \\
  1314.0005  & -18.1  &  2.9 \\
  1340.8393  &  25.2  &  2.4 \\
  1341.8853  &  21.2  &  2.7 \\
  1342.8787  &  25.6  &  2.9 \\
  1367.7877  &  36.4  &  2.8 \\
  1368.7558  &  47.0  &  2.6 \\
  1369.7821  &  43.8  &  3.1 \\
  1370.8677  &  47.2  &  3.1 \\
  1371.7599  &  41.9  &  2.6 \\
  1372.7678  &  46.6  &  2.8 \\
  1373.7712  &  40.6  &  2.7 \\
  1410.7258  &   0.0  &  2.5 \\
  1411.7251  &  -5.5  &  3.0 \\
\enddata
\end{deluxetable}

\clearpage

\begin{deluxetable}{rrr}
\tablenum{6}
\tablecaption{Velocities for HD 177830}
\label{vel177830}
\tablewidth{0pt}
\tablehead{
JD & RV & error \\
(-2450000)   &  (m s$^{-1}$) & (m s$^{-1}$) 
}
\startdata
\tableline
   276.0345  &   0.0  &  2.5 \\
   605.0434  &  13.4  &  2.3 \\
   666.8855  &   0.0  &  2.3 \\
   982.9395  &  18.7  &  2.1 \\
  1009.9321  &   5.3  &  2.7 \\
  1068.8172  & -15.0  &  2.0 \\
  1069.8500  & -14.1  &  1.9 \\
  1070.8953  & -10.5  &  2.2 \\
  1071.8312  & -13.4  &  2.0 \\
  1072.8201  &  -9.7  &  2.1 \\
  1073.8180  & -11.7  &  1.9 \\
  1074.8078  & -17.8  &  2.0 \\
  1075.8981  & -22.4  &  2.6 \\
  1311.1097  &  51.3  &  2.2 \\
  1312.1075  &  46.0  &  2.6 \\
  1313.1058  &  36.9  &  2.2 \\
  1314.1286  &  45.2  &  2.2 \\
  1341.9544  &  50.7  &  2.0 \\
  1367.9145  &  31.6  &  2.6 \\
  1368.9065  &  31.6  &  2.6 \\
  1369.9181  &  24.9  &  2.4 \\
  1409.8468  &  -3.5  &  2.4 \\
  1410.8018  &  -6.1  &  2.4 \\
  1411.7991  &   2.7  &  2.3 \\
  1438.7419  &  -2.1  &  2.0 \\
  1439.7602  &  -6.2  &  2.2 \\
  1440.8698  &  -3.4  &  2.3 \\
  1441.7234  &  -3.5  &  2.2 \\
  1488.7226  &  -3.0  &  2.7 \\
\enddata
\end{deluxetable}

\clearpage

\begin{deluxetable}{rrr}
\tablenum{7}
\tablecaption{Velocities for HD 192263}
\label{vel192263}
\tablewidth{0pt}
\tablehead{
JD & RV & error \\
(-2450000)   &  (m s$^{-1}$) & (m s$^{-1}$) 
}
\startdata
\tableline
   984.0580  &  23.5  &  2.8 \\
  1011.9079  & -21.3  &  2.6 \\
  1050.8738  &  11.6  &  2.6 \\
  1069.8856  & -45.3  &  2.4 \\
  1312.0841  & -70.7  &  3.1 \\
  1313.1102  & -63.1  &  2.7 \\
  1342.0531  &   0.0  &  3.4 \\
  1342.9791  &   9.3  &  5.0 \\
  1367.9100  &  24.4  &  3.8 \\
  1409.9277  & -60.8  &  4.6 \\
  1411.8736  & -30.3  &  3.3 \\
  1438.7651  &  -1.5  &  3.4 \\
  1439.8239  &  16.6  &  3.4 \\
  1440.8836  &  14.9  &  4.2 \\
  1441.8293  &  23.6  &  3.6 \\
\enddata
\end{deluxetable}

\clearpage

\begin{deluxetable}{rrr}
\tablenum{8}
\tablecaption{Velocities for HD 222582}
\label{vel222582}
\tablewidth{0pt}
\tablehead{
JD & RV & error \\
(-2450000)   &  (m s$^{-1}$) & (m s$^{-1}$) 
}
\startdata
\tableline
   805.7232  &  40.0  &  3.3 \\
   984.1216  & -83.5  &  3.9 \\
  1014.1166  &-101.7  &  3.2 \\
  1050.9569  &-118.6  &  3.1 \\
  1051.9943  &-117.5  &  3.1 \\
  1071.9679  &-126.7  &  3.2 \\
  1072.8569  &-130.5  &  2.9 \\
  1173.7117  &-158.4  &  5.0 \\
  1342.0968  & 131.4  &  3.8 \\
  1368.0064  &  69.5  &  3.6 \\
  1369.0076  &  59.4  &  5.0 \\
  1370.0873  &  63.8  &  3.1 \\
  1371.0903  &  59.0  &  3.1 \\
  1372.0672  &  54.5  &  3.8 \\
  1373.0733  &  50.0  &  4.1 \\
  1374.0611  &  52.1  &  3.3 \\
  1410.0295  &   9.8  &  3.2 \\
  1411.0017  &   0.0  &  3.2 \\
  1411.9651  &   0.4  &  3.1 \\
  1438.8624  & -22.9  &  3.5 \\
  1439.9007  & -28.4  &  3.0 \\
  1440.8981  & -25.4  &  2.8 \\
  1441.9259  & -21.6  &  3.4 \\
  1488.7946  & -60.1  &  3.4 \\
\enddata
\end{deluxetable}

\begin{deluxetable}{rrrlllll}
\tablenum{9}
\tablecaption{Stellar Properties of Previously Announced Planets}
\label{candid}
\tablewidth{0pt}
\tablehead{
\colhead{Star} & \colhead{Star} & \colhead{Spec} & \colhead{M$_{\rm Star}$} & \colhead{V} & \colhead{R'$_{\rm HK}$} & \colhead{[Fe/H]} & \colhead{d} 
\\
\colhead{(HD)} & \colhead{(Hipp)} &\colhead{type} & \colhead{(M$_{\odot}$)} & \colhead{(mag)}  & \colhead{ } & \colhead{ } & \colhead{(pc)}
} 
\startdata
 187123 &  97336 & G3 ~V & 1.00 & 7.83 & -5.00 & +0.16 & 47.9 \\
 195019 & 100970 & G3 ~V & 0.98 & 6.87 & -5.02 & +0.00 & 37.4 \\
 210277 & 109378 & G7 ~V & 0.92 & 6.54 & -5.03 & +0.24 & 21.3 \\
 217107 & 113421 & G7 ~V & 0.96 & 6.17 & -5.06 & +0.29 & 19.7 \\

\enddata
\end{deluxetable}

\clearpage

\clearpage

\begin{deluxetable}{rrllllllll}
\tablenum{10}
\tablecaption{Orbital Parameters of Previously Announced Planets}
\label{candid}
\tablewidth{0pt}
\tablehead{
\colhead{Star}  & \colhead{Period} & \colhead{$K$} & \colhead{$e$} & \colhead{$\omega$} & \colhead{$T_0$} & \colhead{\msini} & \colhead{N} &
\colhead{$RMS$} 
\\
\colhead{(HD)} & \colhead{(days)} & \colhead{(\ms)} &\colhead{ } & \colhead{(deg)} & \colhead{(JD-2450000)}  & \colhead{(\mjup)} & \colhead{ } & \colhead{(\ms)}
} 
\startdata
187123\tablenotemark{a} &  3.0966  (0.0002) &  ~68 (2) & 0.01 (0.03) & 324 (45) & 1013.8 (0.31) & 0.48 & 41 & 3.29 \\
195019                  &  18.200 (0.006)  & 272 (4)   & 0.02 (0.02) & 227 (33) & 1306.5 (2) & 3.47 & 14 & 3.46 \\
210277                  &   436.6 (4)      & ~39 (2)   & 0.45 (0.03) & 118 (6) &  1428 (4) & 1.23 & 45 & 3.24 \\
217107\tablenotemark{b} &   7.1260 (0.0007) & 140 (3) & 0.14 (0.02)  & ~31 (7) &  1331.4 (0.2) & 1.27 & 21 & 4.20 \\
\enddata
\tablenotetext{a}{Additional Slope is -7 $\pm$2 \ms per yr.}
\tablenotetext{b}{Additional Slope is 40 $\pm$3 \ms per yr.}
\end{deluxetable}

\clearpage

\begin{deluxetable}{rrr}
\tablenum{11}
\tablecaption{Velocities for HD 187123}
\label{vel187123}
\tablewidth{0pt}
\tablehead{
JD & RV & error \\
(-2450000)   &  (m s$^{-1}$) & (m s$^{-1}$) 
}
\startdata
\tableline
   805.7017  &  -9.1  &  2.3 \\
   983.0277  &  76.9  &  2.6 \\
  1009.9419  & -23.9  &  2.7 \\
  1011.1156  &  74.9  &  3.1 \\
  1011.8744  &  -8.4  &  2.7 \\
  1012.0665  & -31.8  &  2.6 \\
  1012.8400  & -45.2  &  2.9 \\
  1012.9485  & -30.0  &  2.6 \\
  1013.0751  & -18.4  &  2.4 \\
  1013.7906  &  73.0  &  2.8 \\
  1013.9150  &  72.3  &  2.8 \\
  1014.0811  &  80.6  &  2.6 \\
  1043.0075  & -30.2  &  1.8 \\
  1043.9607  & -32.5  &  3.2 \\
  1044.0560  & -23.6  &  2.6 \\
  1050.7296  &  44.5  &  1.7 \\
  1051.0040  &  73.0  &  1.9 \\
  1051.7296  &  42.9  &  1.6 \\
  1052.0122  &   5.5  &  2.1 \\
  1068.8311  & -19.3  &  2.3 \\
  1069.8458  &  71.7  &  2.5 \\
  1070.8914  & -28.1  &  2.5 \\
  1071.8273  & -30.5  &  2.4 \\
  1072.8169  &  74.3  &  3.0 \\
  1073.8431  &  -9.0  &  2.7 \\
  1074.8048  & -43.6  &  2.9 \\
  1312.1143  &   0.0  &  3.1 \\
  1341.0340  & -50.5  &  3.8 \\
  1341.9865  &  57.6  &  2.8 \\
  1342.9757  &  14.5  &  3.1 \\
  1367.9201  &  -6.9  &  3.1 \\
  1368.9109  & -58.9  &  2.9 \\
  1370.0220  &  67.2  &  3.0 \\
  1370.9293  &   7.7  &  2.5 \\
  1372.0223  & -50.3  &  2.9 \\
  1373.0195  &  66.1  &  2.7 \\
  1373.8137  &  42.3  &  3.5 \\
  1409.9445  &  33.2  &  2.9 \\
  1410.9016  &  41.9  &  2.9 \\
  1411.8813  & -67.2  &  3.0 \\
  1439.7663  & -69.3  &  2.7 \\
\enddata
\end{deluxetable}

\clearpage

\begin{deluxetable}{rrr}
\tablenum{12}
\tablecaption{Velocities for HD 195019}
\label{vel195019}
\tablewidth{0pt}
\tablehead{
JD & RV & error \\
(-2450000)   &  (m s$^{-1}$) & (m s$^{-1}$) 
}
\startdata
\tableline
  1068.8520  &-319.2  &  2.9 \\
  1069.8930  &-264.0  &  2.9 \\
  1070.9127  &-185.7  &  2.6 \\
  1071.8489  & -91.8  &  2.8 \\
  1072.8369  &   0.0  &  3.0 \\
  1074.8568  & 149.9  &  2.7 \\
  1075.7970  & 189.1  &  4.2 \\
  1312.0909  & 181.0  &  3.1 \\
  1342.0562  &-315.9  &  3.2 \\
  1367.9120  & 187.3  &  3.4 \\
  1409.9302  &-170.9  &  3.1 \\
  1411.8760  &-315.8  &  2.9 \\
  1438.7673  & 144.9  &  2.8 \\
  1439.8259  & 188.9  &  3.1 \\
\enddata
\end{deluxetable}

\clearpage

\begin{deluxetable}{rrr}
\tablenum{13}
\tablecaption{Velocities for HD 210277}
\label{vel210277}
\tablewidth{0pt}
\tablehead{
JD & RV & error \\
(-2450000)   &  (m s$^{-1}$) & (m s$^{-1}$) 
}
\startdata
\tableline
   277.0413  &  -2.2  &  2.6 \\
   366.7926  &  24.6  &  2.2 \\
   462.7062  &  33.5  &  2.6 \\
   605.0940  & -26.3  &  2.2 \\
   665.9876  & -10.4  &  2.6 \\
   688.9457  &  -1.7  &  2.4 \\
   689.9833  &  -1.8  &  2.3 \\
   713.8792  &   3.9  &  2.2 \\
   714.9728  &   3.3  &  2.1 \\
   715.9286  &   9.7  &  2.2 \\
   783.7124  &  12.0  &  2.6 \\
   784.7205  &  26.1  &  2.5 \\
   785.6995  &  19.5  &  2.6 \\
   805.7146  &  16.9  &  2.5 \\
   806.7032  &  23.8  &  1.8 \\
   956.0877  &  24.6  &  2.2 \\
   983.0511  &  -6.8  &  2.7 \\
   984.0878  &  -4.4  &  3.2 \\
  1010.0261  & -39.2  &  2.5 \\
  1011.1015  & -39.4  &  2.2 \\
  1011.9692  & -38.9  &  2.2 \\
  1013.0816  & -40.0  &  2.2 \\
  1014.0859  & -38.3  &  2.7 \\
  1043.0057  & -37.6  &  2.4 \\
  1043.9942  & -34.4  &  2.5 \\
  1050.9169  & -35.3  &  1.6 \\
  1051.9839  & -33.8  &  2.0 \\
  1068.8670  & -24.6  &  2.1 \\
  1069.9748  & -24.9  &  2.2 \\
  1070.9566  & -21.8  &  2.3 \\
  1071.8706  & -23.6  &  2.3 \\
  1072.9307  & -27.0  &  2.4 \\
  1074.8716  & -19.4  &  2.3 \\
  1170.6885  &   6.1  &  2.2 \\
  1172.6894  &   6.1  &  2.5 \\
  1173.6868  &   6.4  &  2.1 \\
  1311.1031  &  32.2  &  2.3 \\
  1312.0938  &  33.7  &  2.1 \\
  1313.1131  &  39.0  &  2.5 \\
  1341.1060  &  38.5  &  2.7 \\
  1342.0662  &  38.9  &  2.0 \\
  1367.9567  &  35.9  &  2.5 \\
  1410.0179  &   5.0  &  2.4 \\
  1438.8105  & -35.7  &  2.5 \\
  1441.9290  & -39.9  &  2.4 \\
\enddata
\end{deluxetable}

\clearpage

\begin{deluxetable}{rrr}
\tablenum{14}
\tablecaption{Velocities for HD 217107}
\label{vel217107}
\tablewidth{0pt}
\tablehead{
JD & RV & error \\
(-2450000)   &  (m s$^{-1}$) & (m s$^{-1}$) 
}
\startdata
\tableline
  1068.8595  & -20.9  &  2.7 \\
  1069.9728  &-114.4  &  2.9 \\
  1070.9542  &-119.6  &  2.0 \\
  1071.8687  & -68.2  &  2.4 \\
  1072.9288  &  34.5  &  2.7 \\
  1074.8695  & 128.6  &  2.5 \\
  1075.8261  &  -8.1  &  3.6 \\
  1171.7040  & -58.4  &  4.6 \\
  1172.7064  &  53.3  &  4.9 \\
  1173.7052  & 142.3  &  4.8 \\
  1312.1031  & -78.6  &  3.1 \\
  1343.0337  &  -2.9  &  2.9 \\
  1367.9608  &  37.2  &  3.5 \\
  1371.0873  & -43.4  &  2.7 \\
  1372.0642  &  50.7  &  2.8 \\
  1373.0712  & 161.3  &  3.2 \\
  1374.0587  & 169.1  &  3.2 \\
  1410.0248  & 140.3  &  2.7 \\
  1410.9805  &   0.0  &  3.0 \\
  1411.9452  & -85.9  &  2.8 \\
  1438.8143  & 107.0  &  2.6 \\
\enddata
\end{deluxetable}

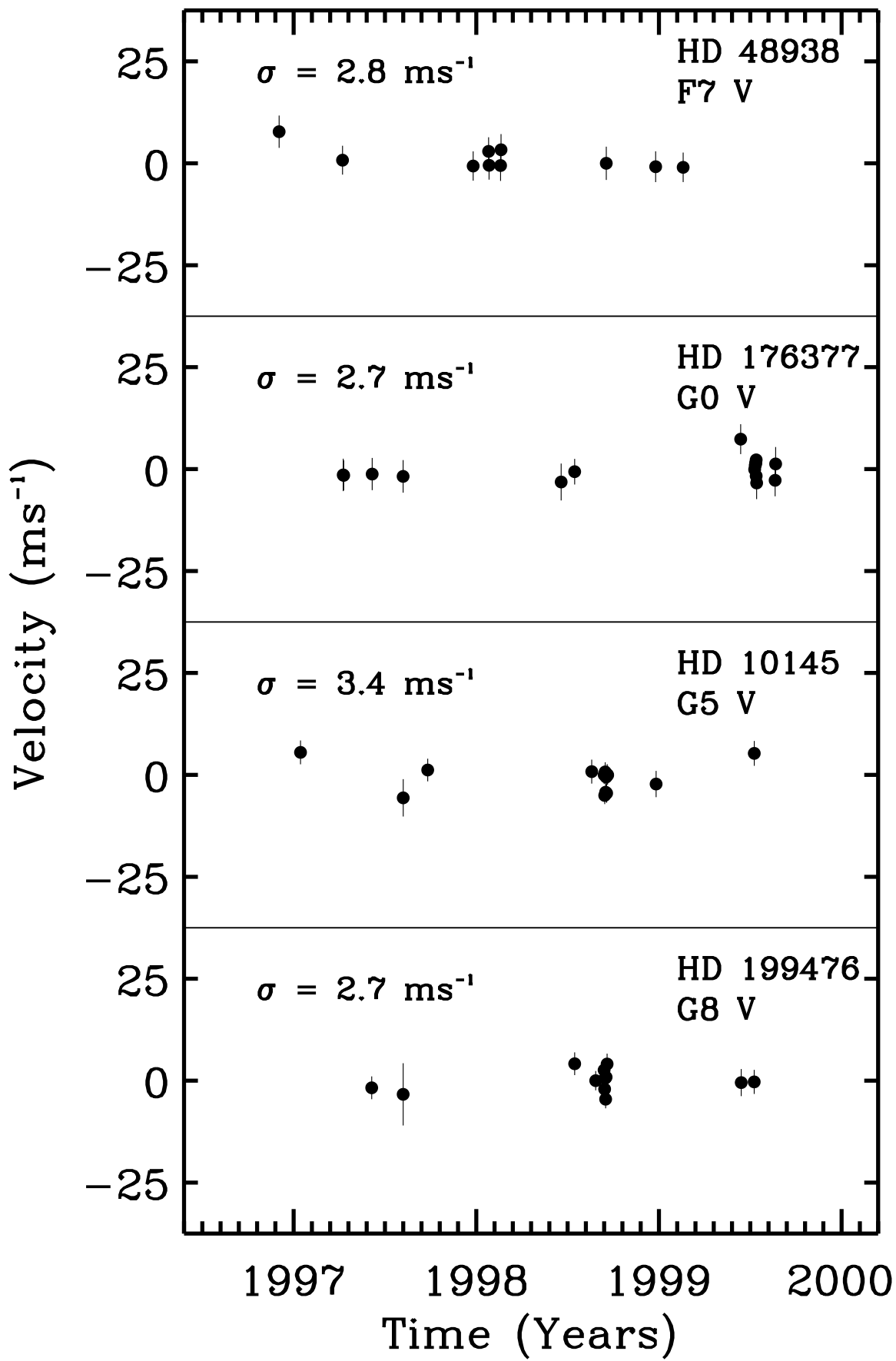
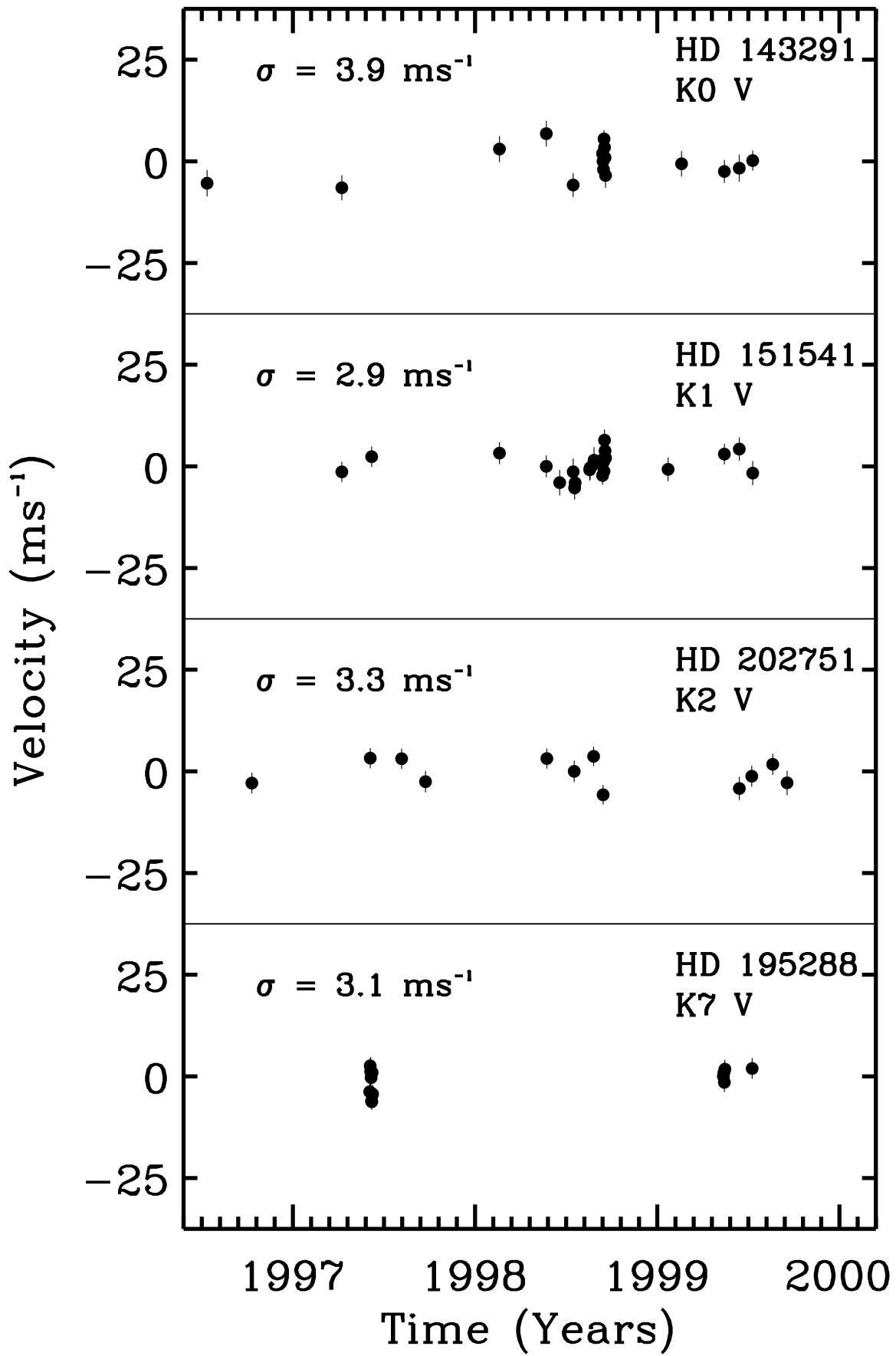
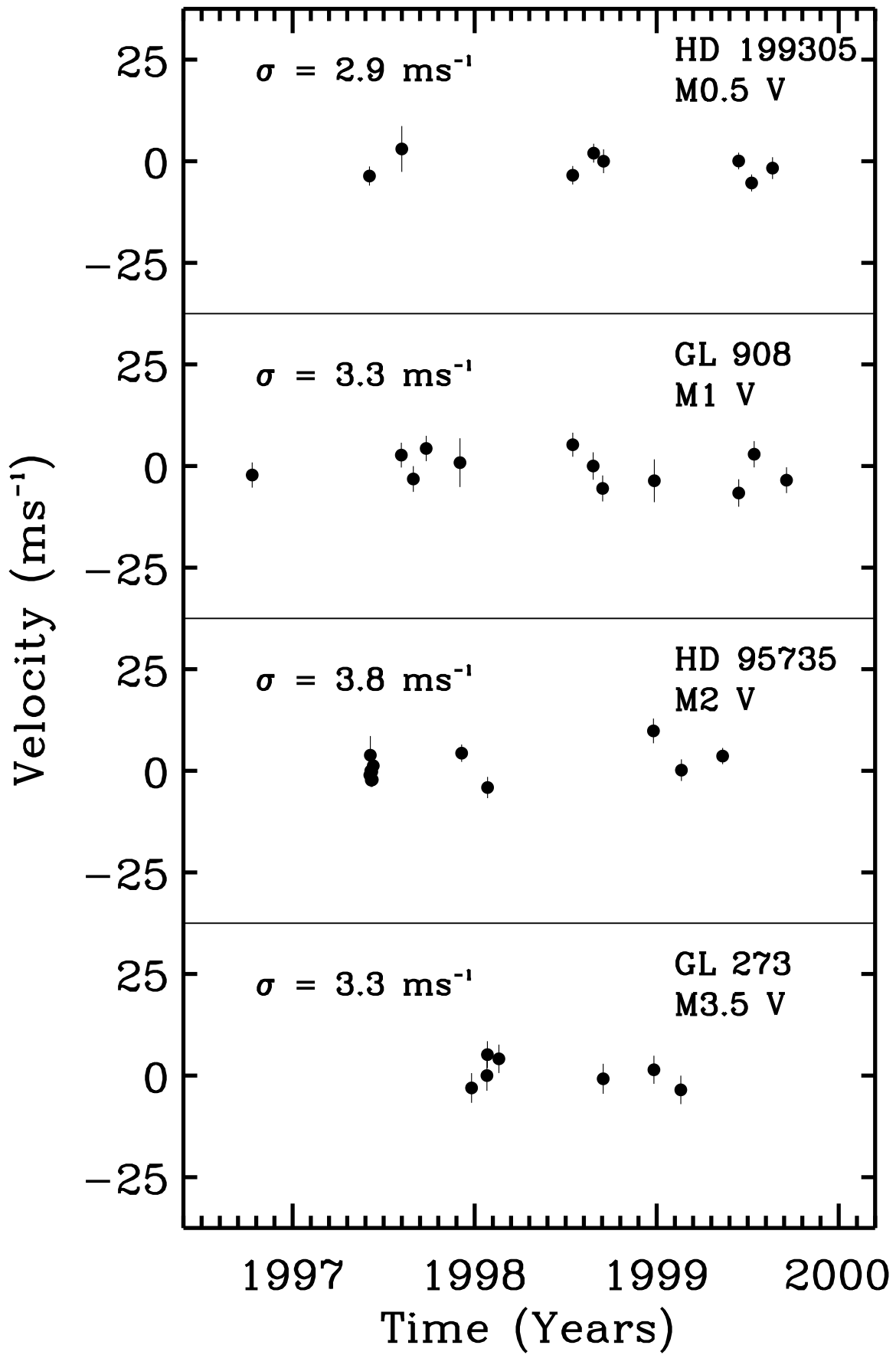
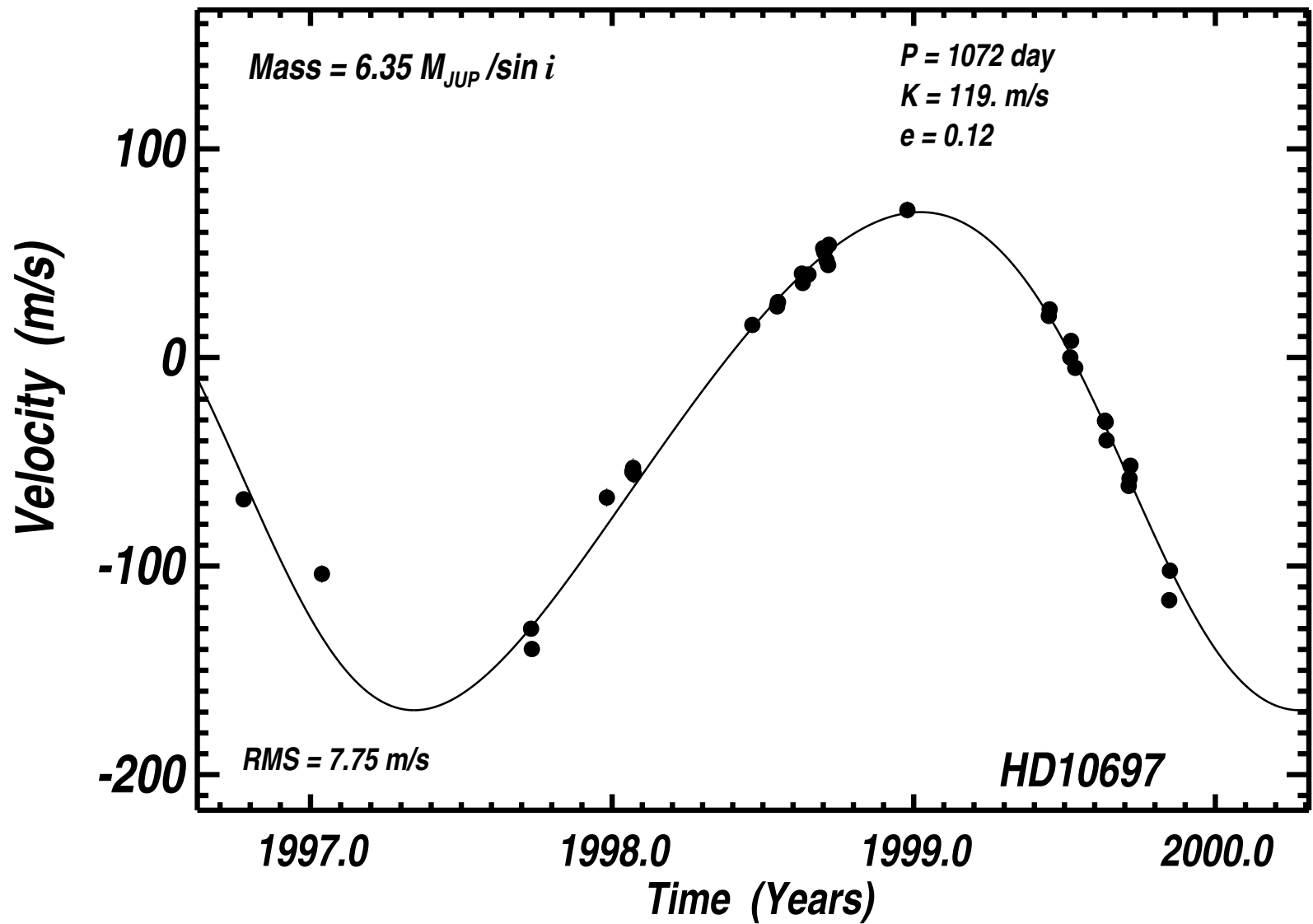
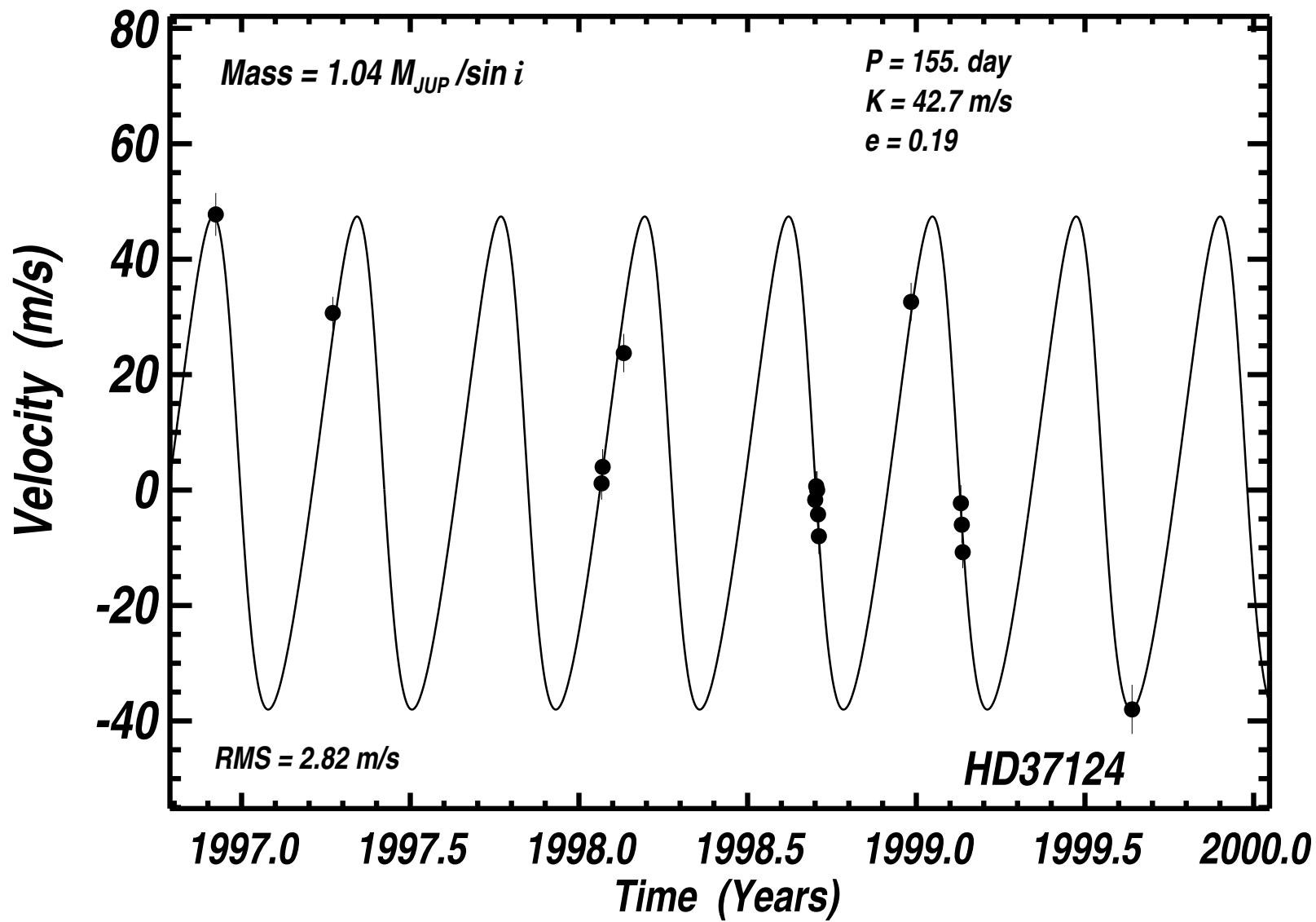
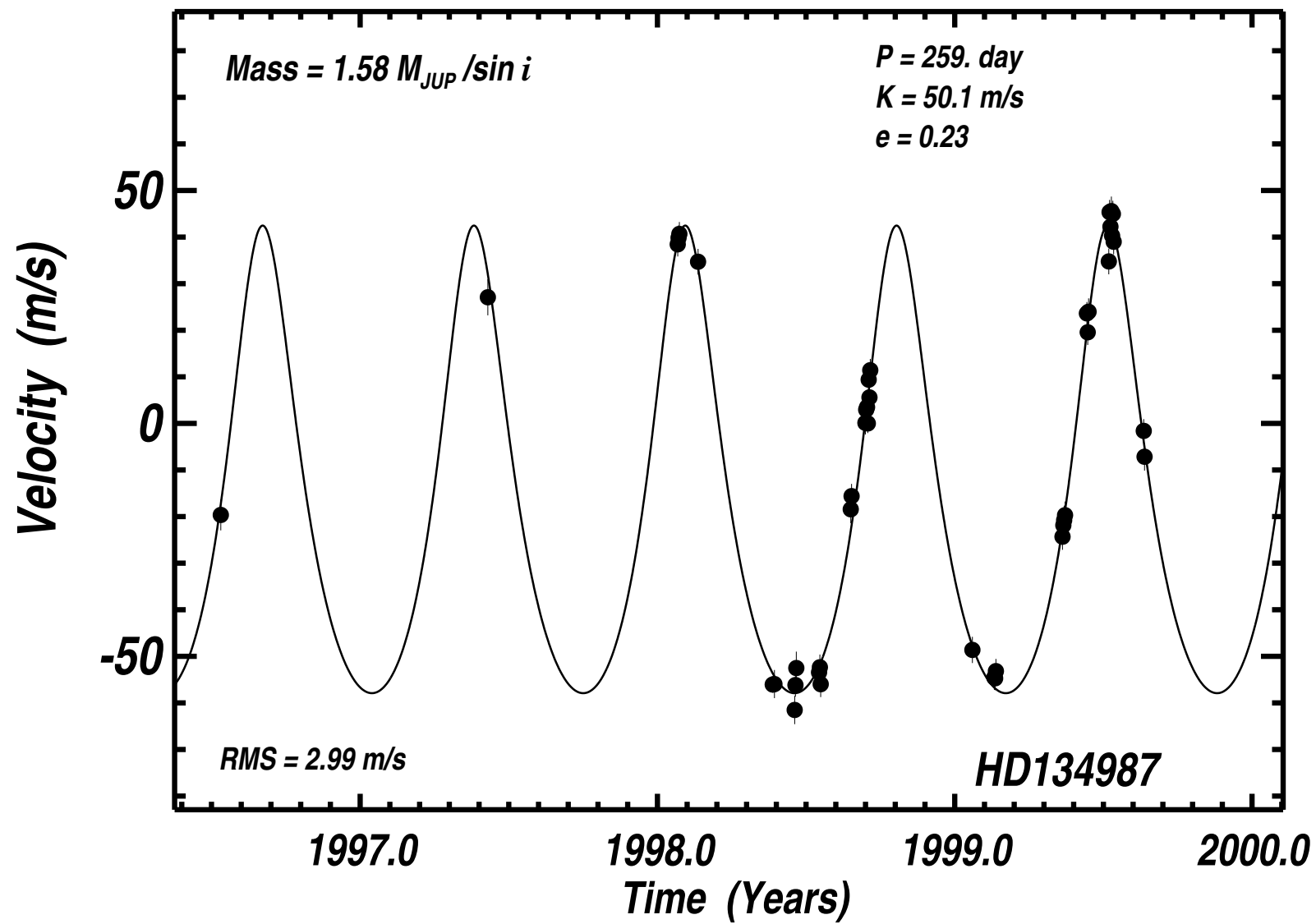
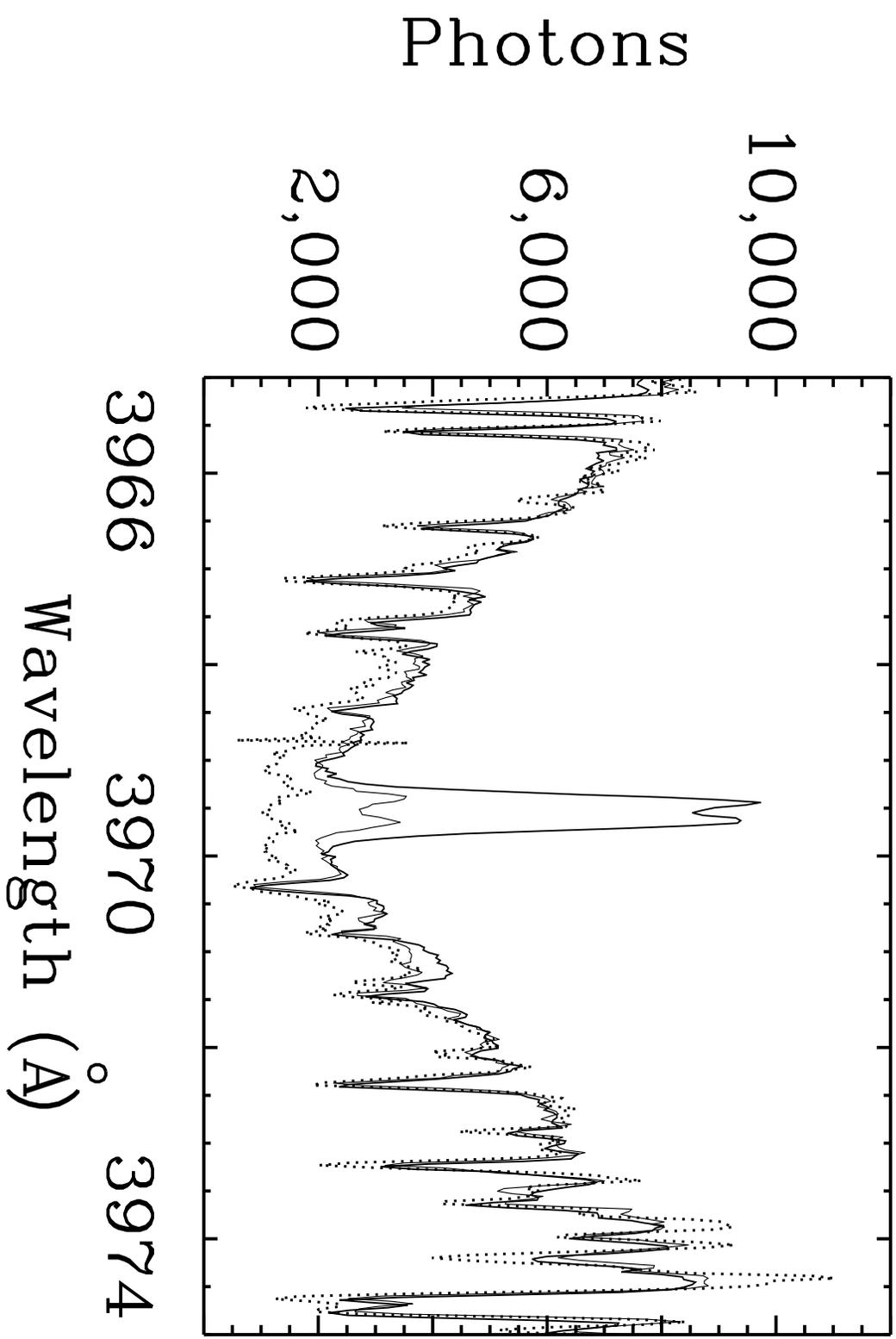
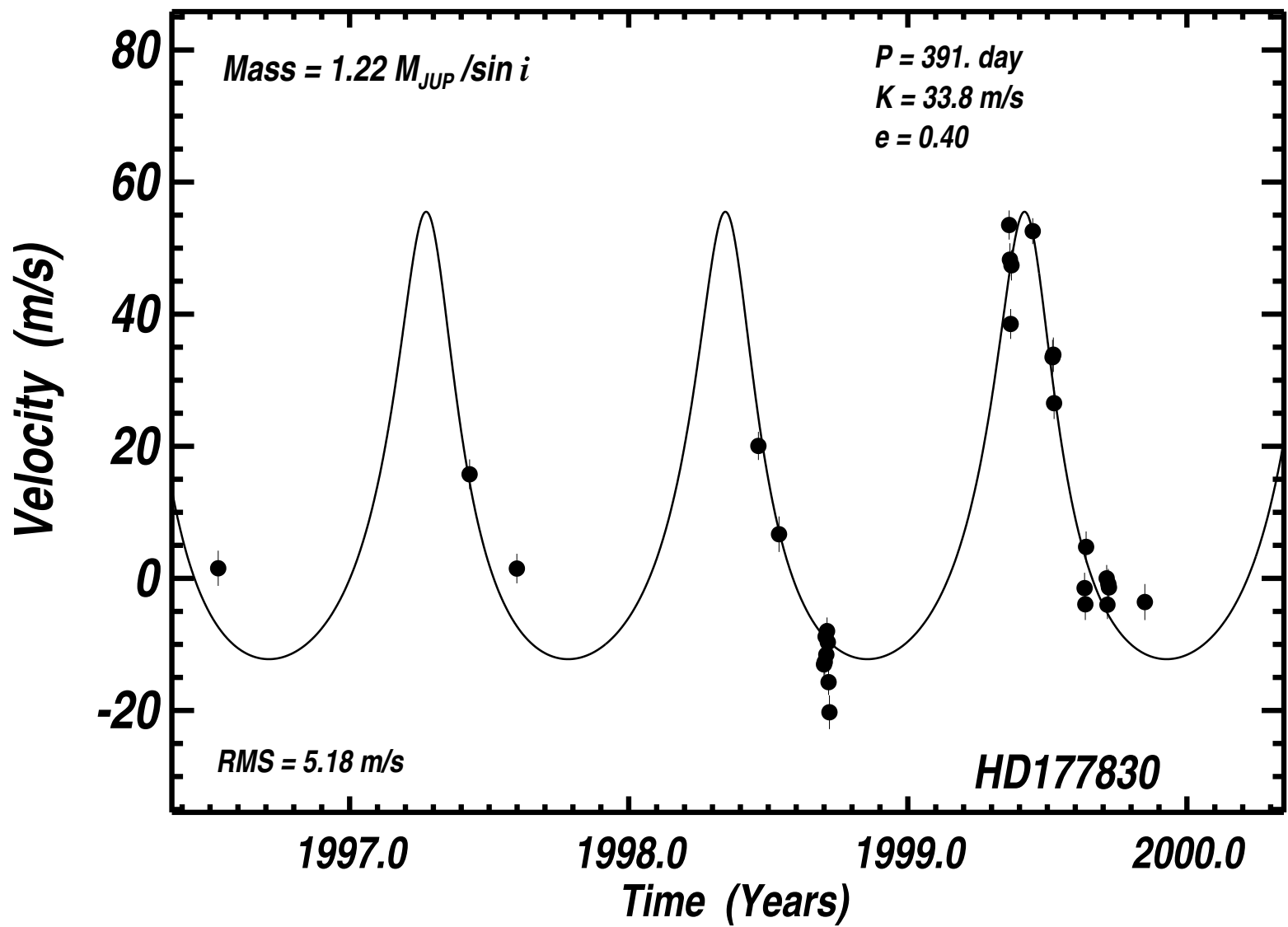
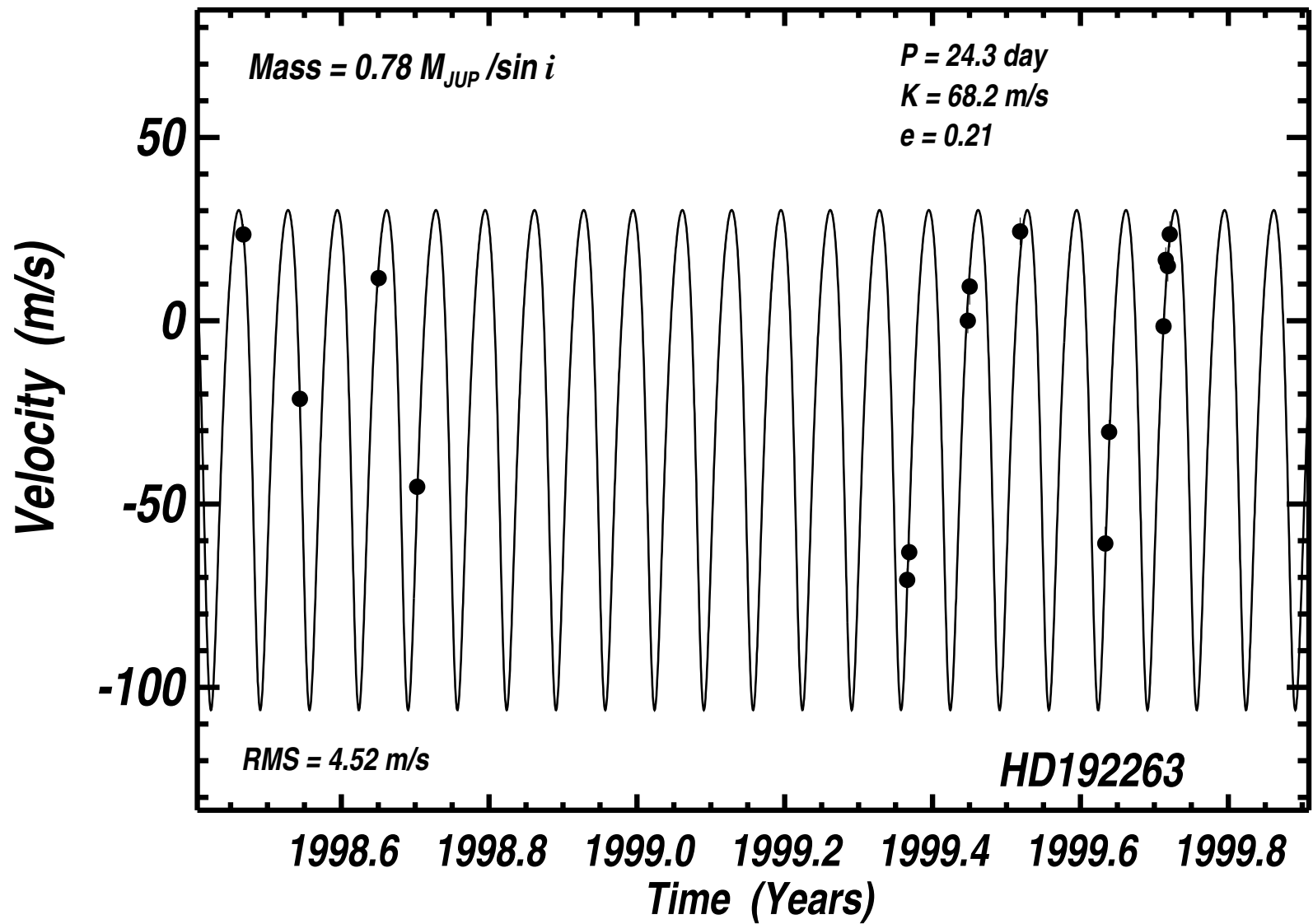
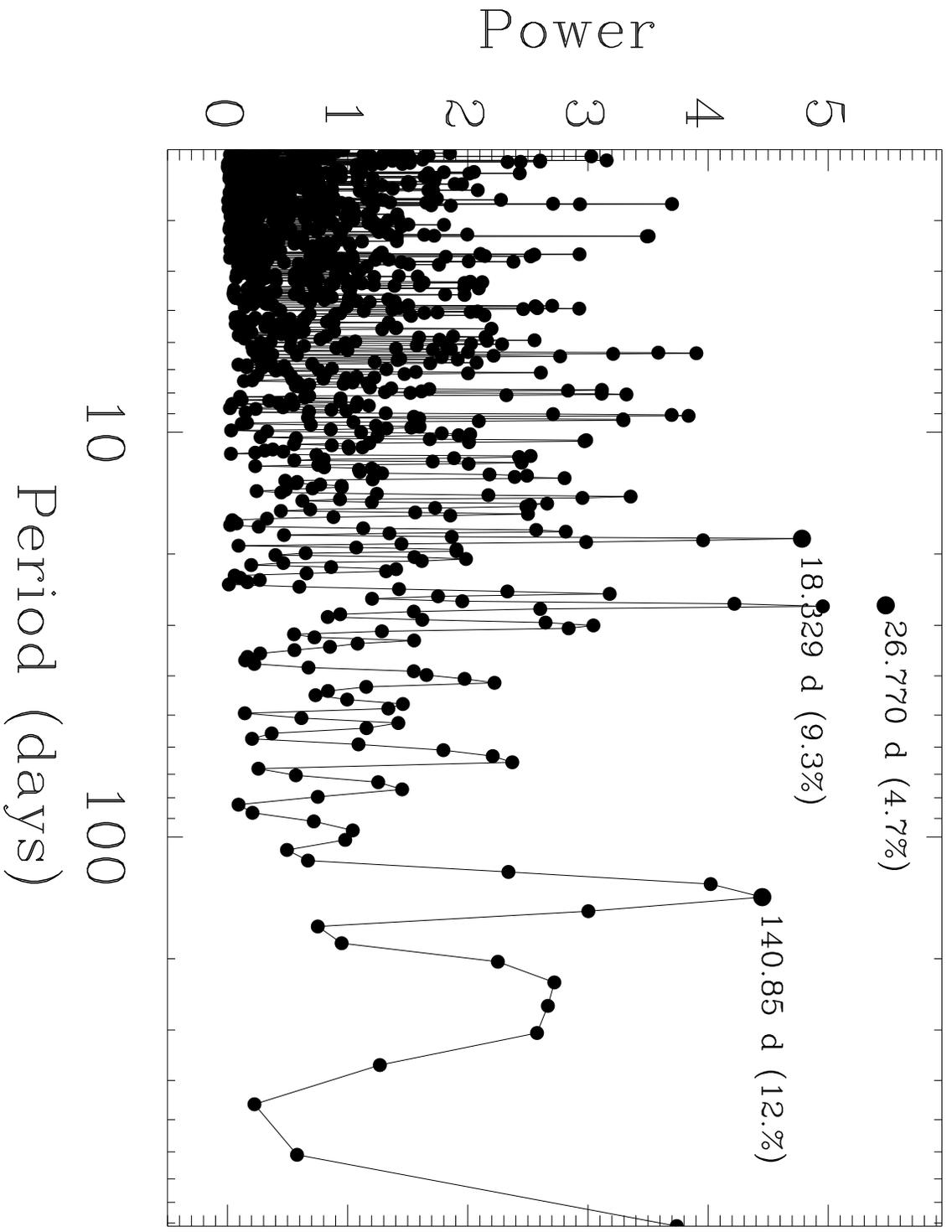
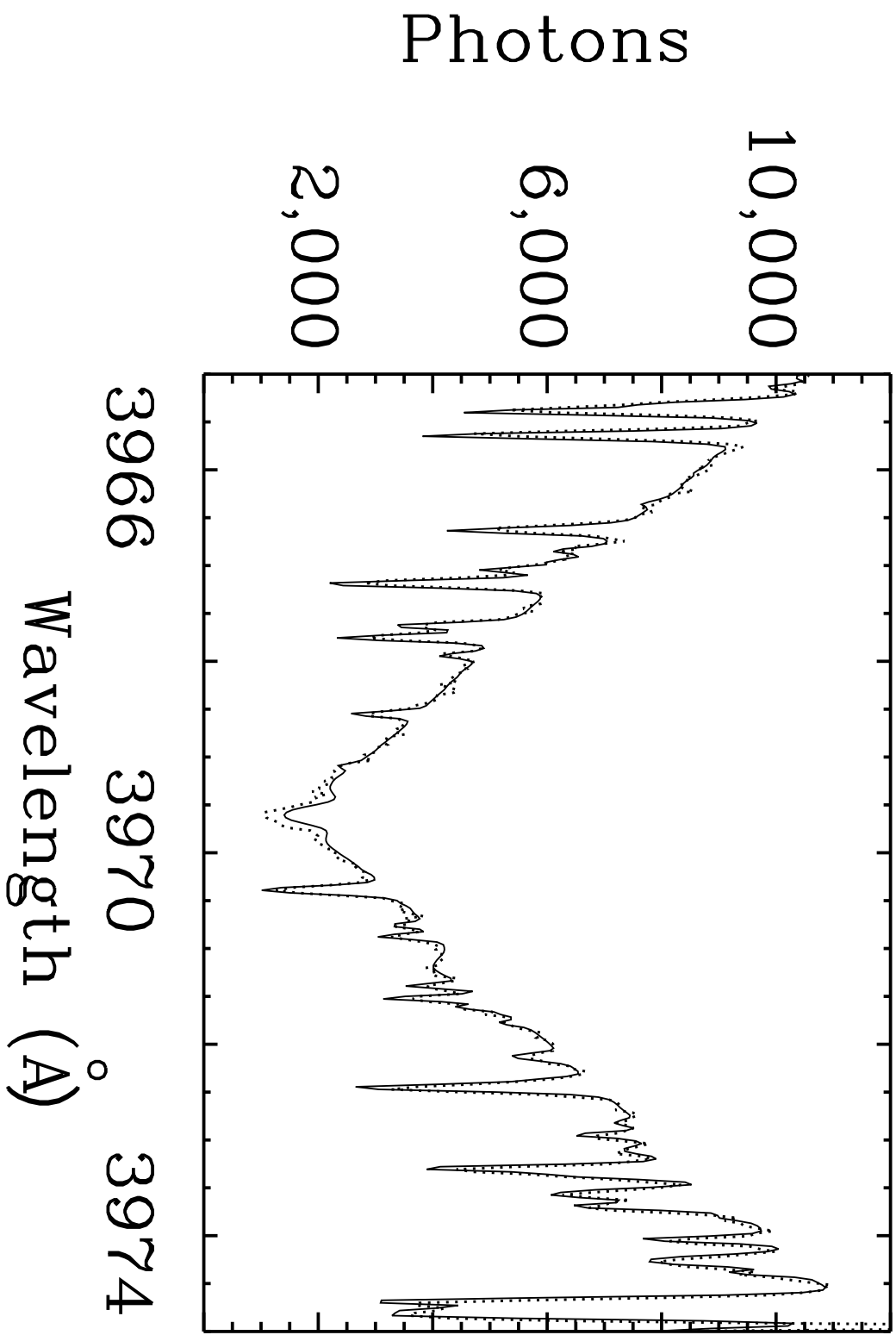
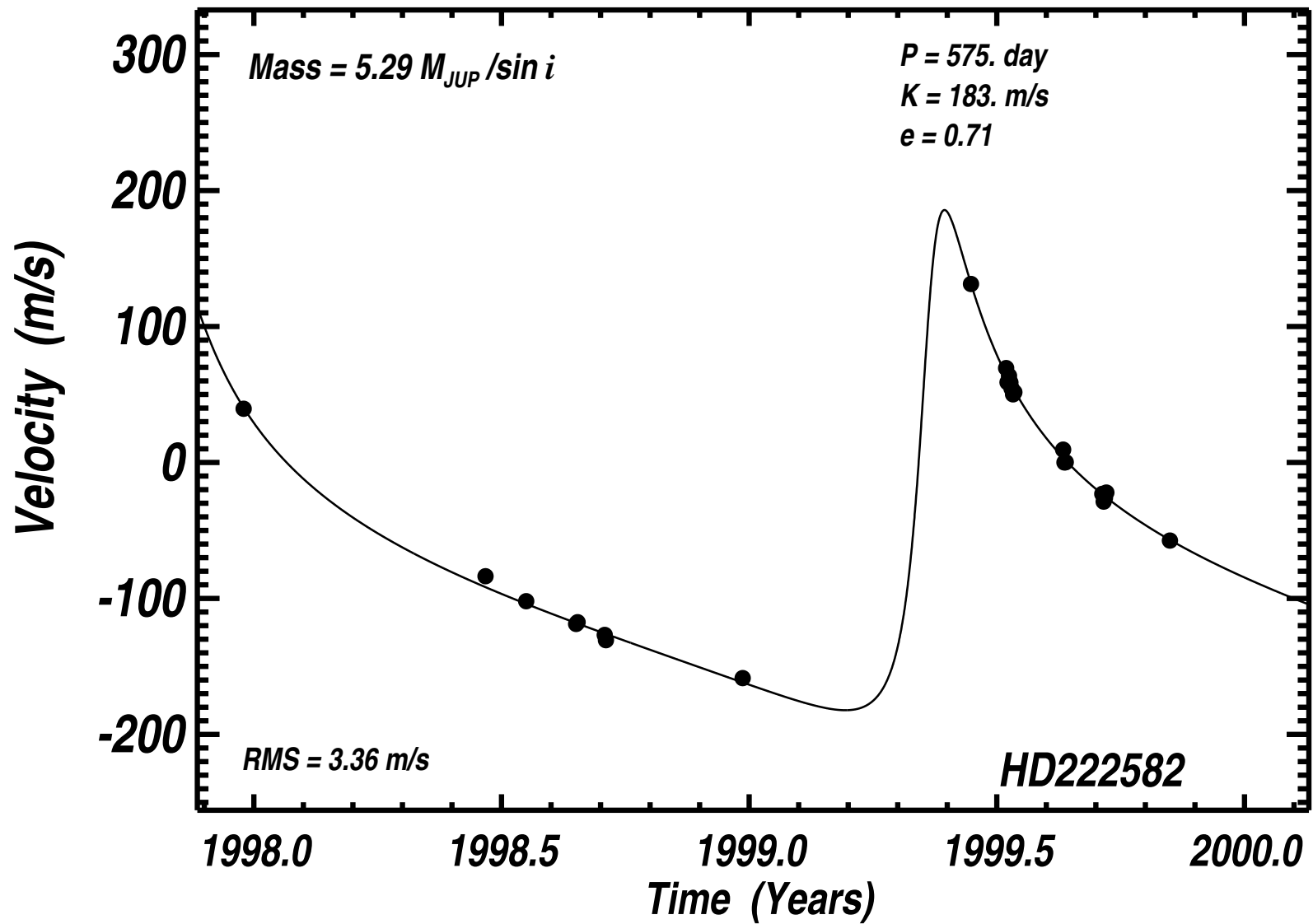
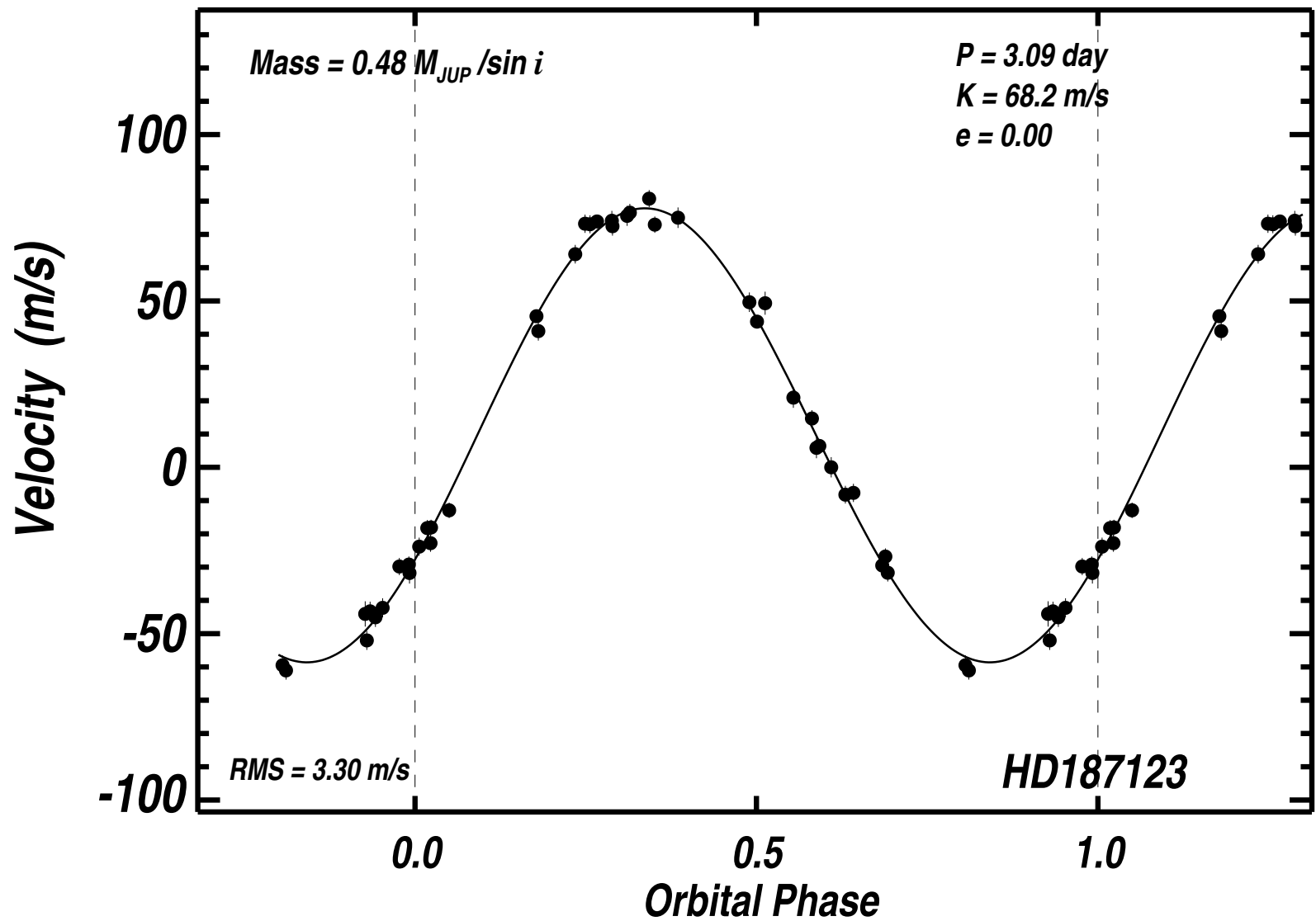
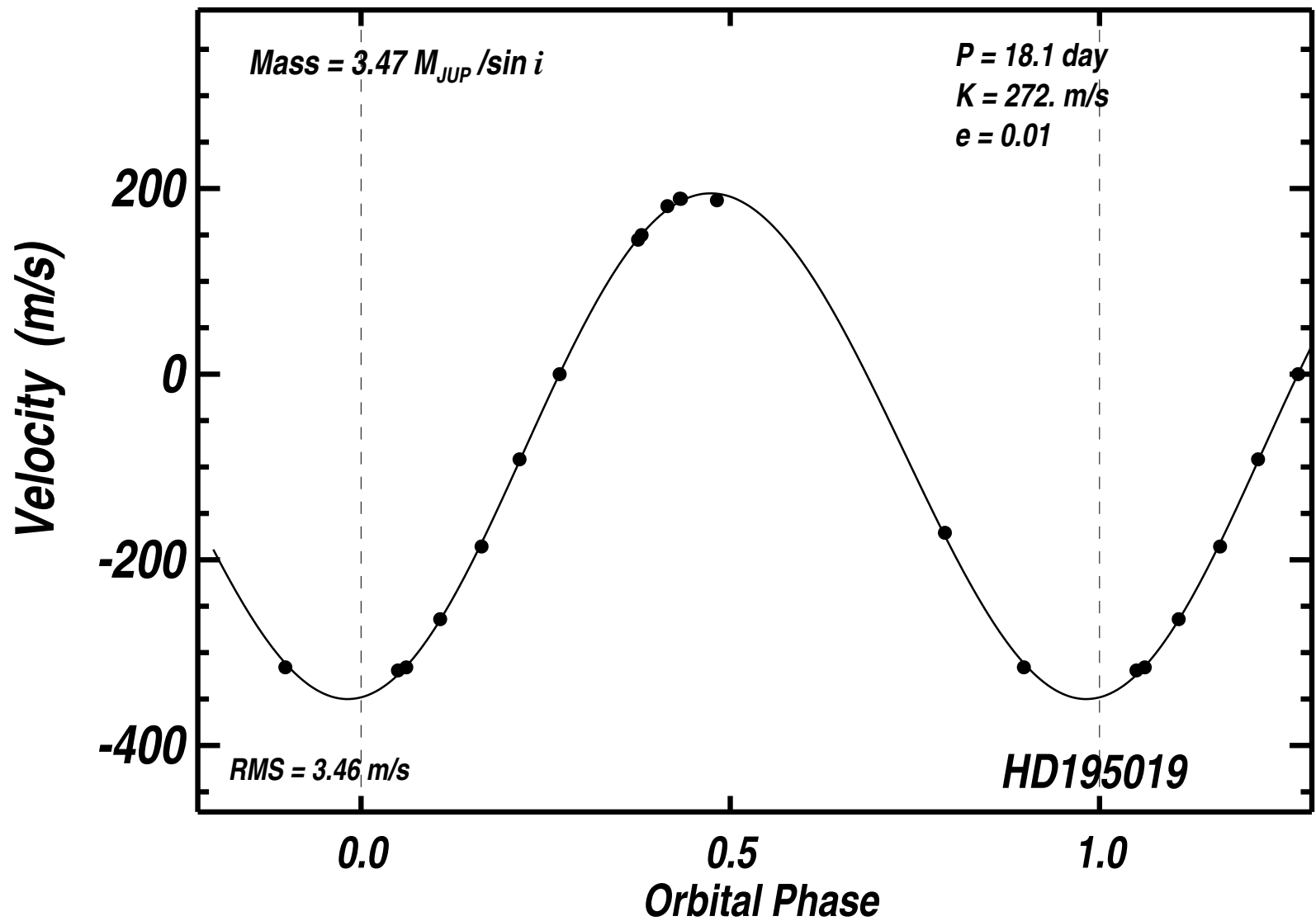
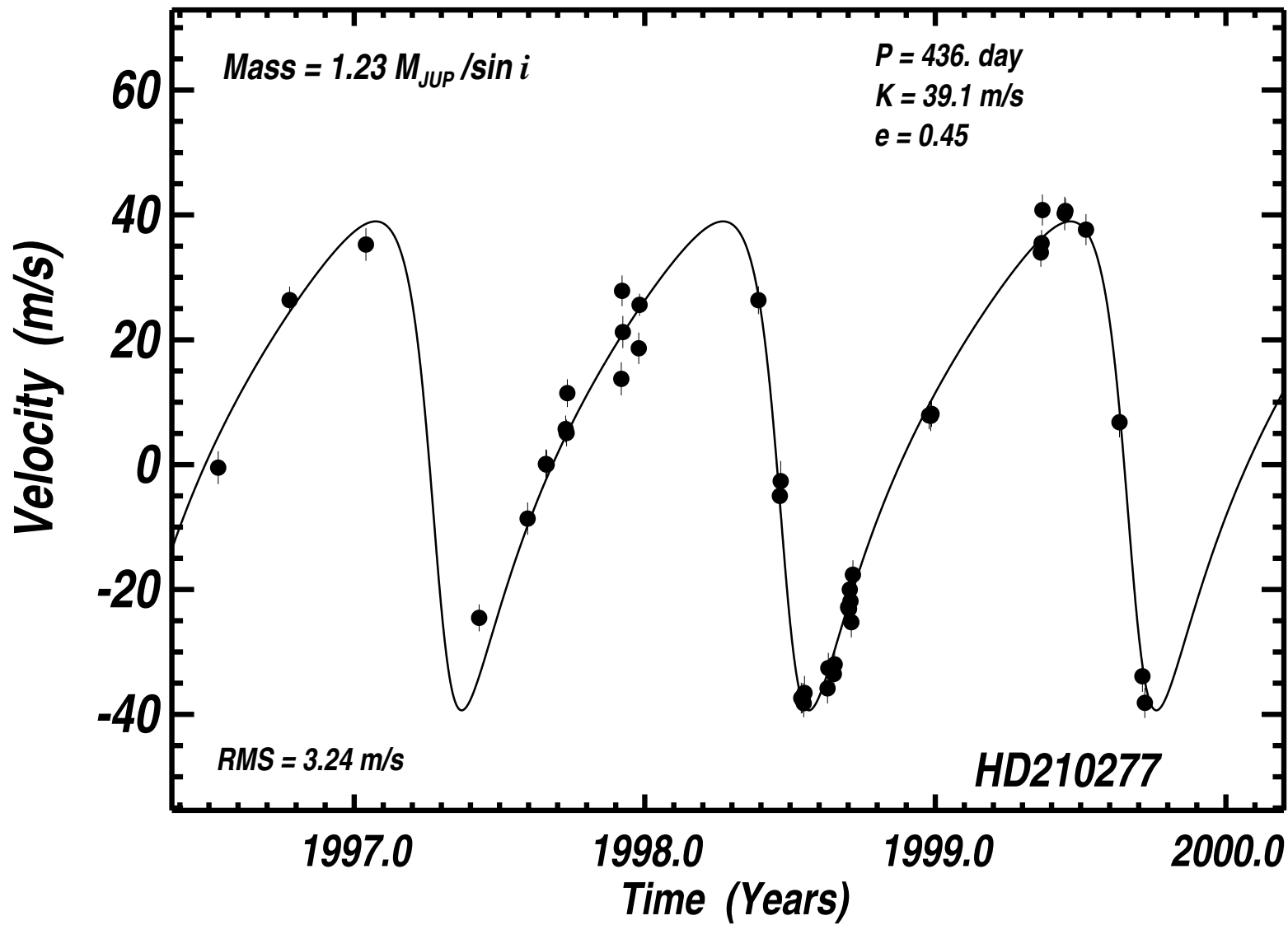
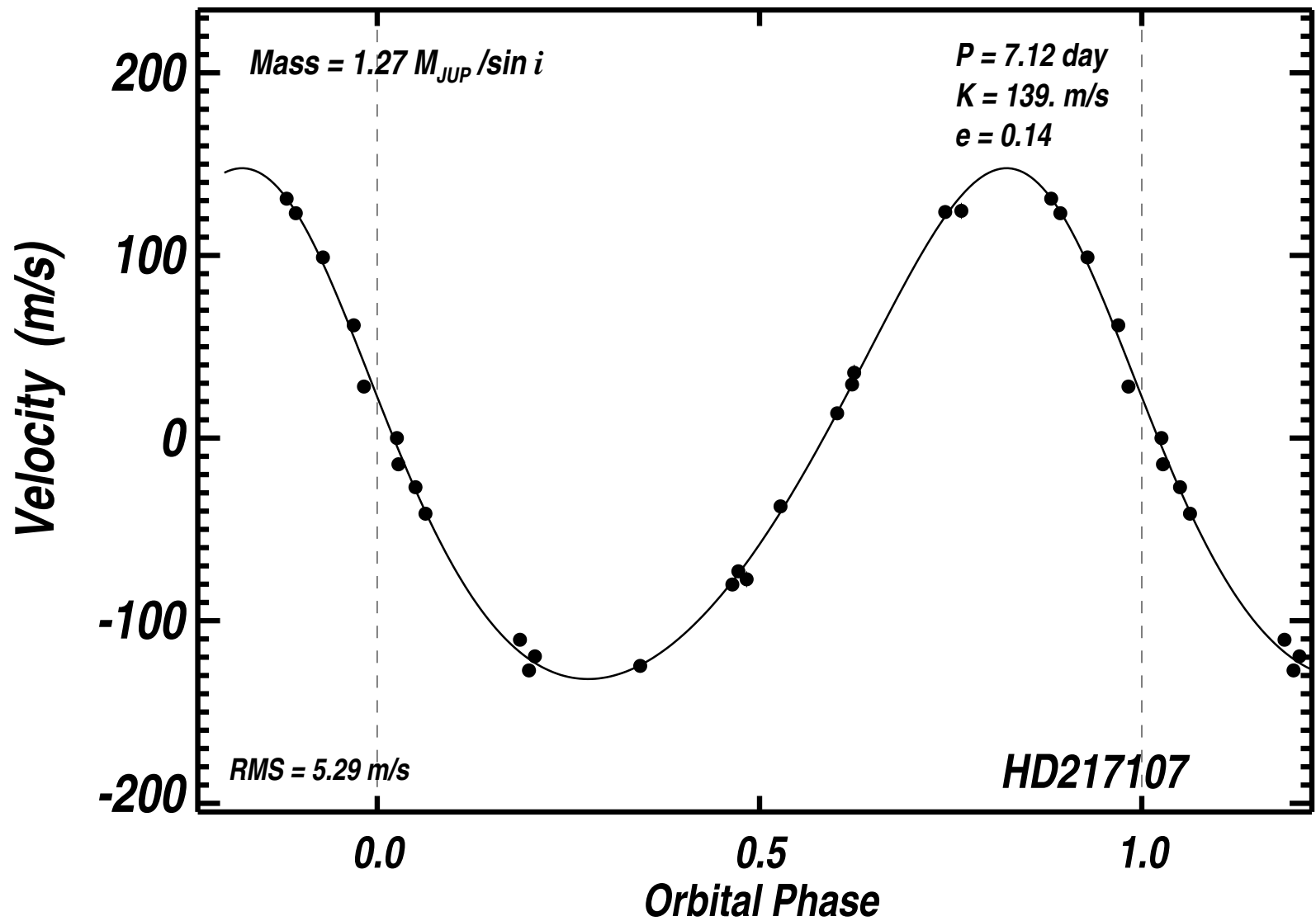
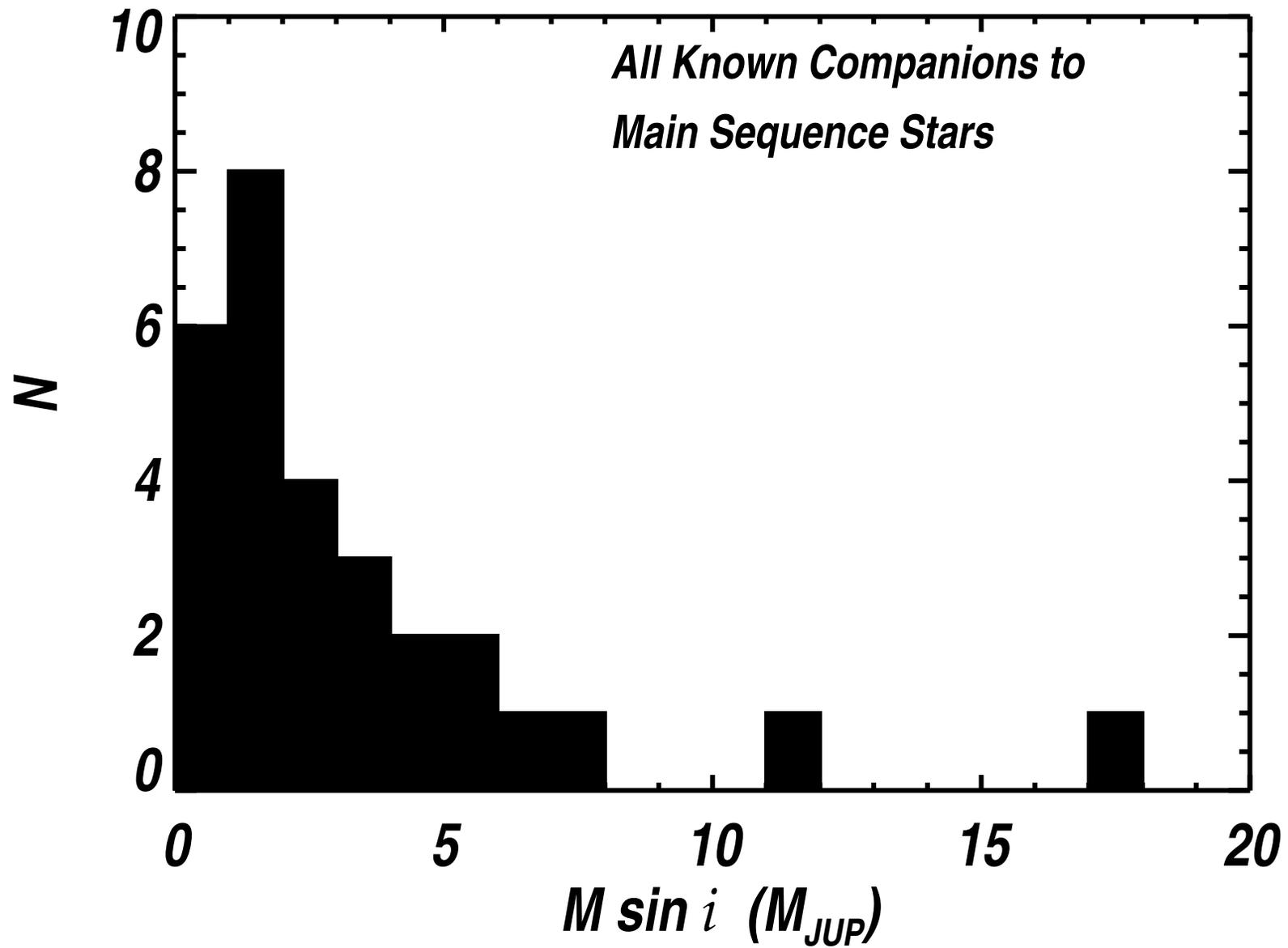
\end{document}